\documentclass[aps,prm,reprint,showpacs,showkeys,superscriptaddress,floatfix,twocolumn]{revtex4-2}
\usepackage{multirow}
\usepackage{graphicx}
\usepackage{bm}
\usepackage{dcolumn}
\usepackage[colorlinks=true,linkcolor=blue,urlcolor=blue,citecolor=blue]{hyperref}
\usepackage{natbib}
\usepackage{amsmath}
\usepackage{times}
\usepackage{xcolor}
\usepackage{soul}
\usepackage[normalem]{ulem}

\begin{document}

\title{Realization of two-sublattice exchange physics in the triangular lattice compound Ba$_3$Er(BO$_3$)$_3$}

\author{Matthew Ennis}\affiliation{Department of Physics, Duke University, Durham, NC, USA}
\author{Rabindranath Bag}\affiliation{Department of Physics, Duke University, Durham, NC, USA}
\author{Chunxiao Liu}\affiliation{Department of Physics, University of California,  Berkeley, CA, USA}
\author{Sachith E. Dissanayake}\affiliation{Department of Physics, Duke University, Durham, NC, USA}
\author{Alexander I. Kolesnikov}\affiliation{Neutron Scattering Division, Oak Ridge National Laboratory, Oak Ridge, TN, USA}
\author{Leon Balents}\affiliation{Kavli Institute for Theoretical Physics, University of California, Santa Barbara, CA, USA}
\author{Sara Haravifard}
\email[email:]{sara.haravifard@duke.edu} \affiliation{Department of Physics, Duke University, Durham, NC, USA} \affiliation{Department of Mechanical Engineering and Materials Science, Duke University, Durham, NC, USA}

\date{\today}

\begin{abstract}
We have synthesized high-quality single crystal samples of the erbium-based triangular lattice compound Ba$_3$Er(BO$_3$)$_3$. Thermal and magnetic measurements reveal large anisotropy and a possible phase transition at 100 mK. The low-temperature magnetic heat capacity can be understood from the two Wyckoff positions that the magnetic ions Er$^{3+}$ occupy, which have distinct symmetry properties and crystal field environments. A point charge calculation for the crystal electric field levels is consistent with this understanding. Based on symmetry analysis and classical simulation, we argue that Ba$_3$Er(BO$_3$)$_3$ realizes an interesting two-sublattice exchange physics, in which the honeycomb lattice spins develop ferromagnetic correlations due to the additional spins at the hexagon centers but eventually order antiferromagnetically. Additionally, our results suggest that quantum fluctuations need to be considered in order to fully explain the experimental observations.
\end{abstract}
	
\maketitle


\section{Introduction}
In recent years, materials that exhibit geometric frustration have received much attention for their potential to host a quantum spin liquid (QSL) state \cite{AndersonMatRes1973, BalentsNature2010}. Frustration is commonly realized in materials with magnetic rare-earth ions sitting on a two-dimensional triangular lattice. Of these, YbMgGaO$_4$ has received considerable attention as being a potential host of a quantum spin liquid state \cite{LiSciR2015, LiPRL2015, Li_PRL2016, ZhangPRX2018, Li_YMGOreview}. However, YbMgGaO$_4$ has random site mixing between the Mg and Ga atoms, which can make interpretation of the measurements difficult, and has been the source of much debate about the true nature of the ground state \cite{MaPRL2018, LiPRL2017, ZhuPRL2017, ParkerPRB2018}. This has led to further study of other ytterbium-based triangular materials without site disorder, such as the rare-earth chalcogenides NaYbX$_2$ (X = O, S, Se) \cite{RanjithPRB2019, BordelonNature2019, DingPRB2019, BaenitzPRB2018, GuoPRM2020, LiuChinLett2018, BordelonPRB2020} or the orthoborate materials (Na, K)(Sr, Ba)Yb(BO$_3$)$_2$ \cite{Guo_2019, Sanders_2017, Svetlyakova2013, Pan_PRB, Kuznetsov_NSYBO}. We have recently reported the single crystal study of another triangular lattice borate compound without site mixing: Ba$_3$Yb(BO$_3$)$_3$, and have found it to exhibit a spin-1/2 quantum dipole ground state \cite{BYBOPRB}.

Due to the chemical similarity of the rare-earth elements, different rare-earth elements can be substituted for Yb in the lattice, changing the spin of the magnetic ion without significantly altering the crystal structure. Many of the extensively studied Yb-based  triangular materials have also been reported with other rare-earths substituted into the lattice. Some examples are ErMgGaO$_4$ \cite{Cevallos_SSC2018} and TmMgGaO4 \cite{Cevallos_TmMgGaO4}, the rare-earth chalcogenides ARECh$_2$ (A = Na, K, Rb; RE= rare-earth; Ch = O, S, Se)\cite{Hashimoto_2003, LiuChinLett2018, XingPRM2019, Deng_RbLnSe2,Sanjeewa_JSSC2022, Havlak_2015, Dong_2008} and several examples in the orthoborate family (Na, K)(Sr, Ba)RE(BO$_3$)$_2$ (RE = rare-earth) \cite{Kononova_2016, Sanders_2017, Kuznetsov_NSYBO, Guo_MRE2019, Guo_InorgChem}. 

A particularly interesting case is substituting spin-1/2 Yb with spin-3/2 Er. Since Er$^{3+}$ is also a Kramers ion, it is expected to have a doublet ground state like Yb, so these compounds would provide an opportunity to study the effect of higher spin on the ground states. Both powder and single crystal samples of ErMgGaO$_4$ have been produced \cite{Cevallos_SSC2018} which show similar site mixing to YbMgGaO$_4$. $\mu$SR measurements show no signs of ordering down to 25 mK in ErMgGaO$_4$ and rule out the presence of a glassy state \cite{Cai_muSR2020}. Inelastic neutron scattering studies have been performed on NaErS$_2$ and (K, Cs)ErSe$_2$ to extract the crystal electric field (CEF) parameters; these materials show a large $J_z = \pm 1/2$ component to the ground state, which allows for quantum transitions between the doublet states \cite{Gao_2020, Scheie_2020}. KErSe$_2$ was also found to order into a stripe antiferromagnetic state at low temperatures \cite{Xing_2021} and to host low energy spin-wave-like excitations \cite{Ding_2023}. Motivated by these reported results and our recent work on Ba$_3$Yb(BO$_3$)$_3$ \cite{BYBOPRB}, we have begun studies of the analogous compound Ba$_3$Er(BO$_3$)$_3$.

Here, we report the powder and single crystal synthesis of Ba$_3$Er(BO$_3$)$_3$ as well as magnetic and thermal characterization, and inelastic neutron scattering measurements on this material. We have used these measurements to calculate the CEF spectrum and provide a candidate for the magnetic ground state of Ba$_3$Er(BO$_3$)$_3$. Low temperature heat capacity study on Ba$_3$Er(BO$_3$)$_3$ single crystals reveals a possible phase transition at 100 mK. Based on symmetry analysis and classical simulation we propose that Ba$_3$Er(BO$_3$)$_3$ realizes interesting two-sublattice exchange physics, in which the honeycomb lattice spins develop ferromagnetic correlations due to the additional spins at the hexagon centers but eventually order in an antiferromagnetic phase. 

\section{Experimental Methods}
\label{EXP}
Polycrystalline Ba$_3$Er(BO$_3$)$_3$ samples were synthesized using the solid-state reaction technique. BaCO$_3$ (99.95\% metal basis, Alfa Aesar), H$_3{}^{11}$BO$_3$ ($>$ 99 atom \%, Sigma Aldrich), and  Er$_2$O$_3$ (99.95\%, Alfa Aesar) were used as starting precursor materials. The ${}^{11}\rm{B}$ isotope of boron was used in the synthesis to allow for neutron measurements with these samples, due to its lower neutron absorption cross section compared to ${}^{10}\rm{B}$. Using stoichiometric ratios of the starting materials resulted in some leftover Er$_2$O$_3$ after sintering, so 10\% excess of H$_3$BO$_3$ and BaCO$_3$ by weight were added to account for their loss during synthesis. These were mixed thoroughly in a mortar and pestle and then pressed into a pellet. The pellets were sintered at 500 ${}^{\circ}$C for 12 hours to remove any water the powders could have absorbed, and then at 1000 ${}^{\circ}$C for 48 hours to complete the reaction. As needed, samples were resintered at 1000 ${}^{\circ}$C with intermediate grindings to achieve the pure phase. Once the pure phase was obtained, the powders were pressed into a cylindrical rods using a hydrostatic press under 700 bar pressure, in preparation for single crystal growths. The cylindrical rods (known as feed and seed) were sintered at 1130 ${}^{\circ}$C under O$_2$ atmosphere in a vertical tube furnace to achieve higher density. These rods were then grown into single crystals using a four-mirror optical floating zone furnace. Growth parameters were refined over the course of several growth attempts and eventually high purity centimeter-sized crystals were grown. Magnetic susceptibility and magnetization were measured from 300 K down to 300 mK using a superconducting quantum interference device (SQUID) magnetometer with ${}^3$He attachment. Further measurements were carried out using a Quantum Design Dynacool Physical Property Measurement System (QD PPMS). Heat capacity data were taken from room temperature down to 60 mK using helium-4 and dilution refrigerator (DR) options attached to PPMS. Inelastic neutron scattering (INS) experiments were performed on a powder sample of Ba$_3$Er(BO$_3$)$_3$ on the SEQUOIA spectrometer at Oak Ridge National Laboratory \cite{SEQUOIA} to determine the crystal electric field (CEF) ground state of the compound. Measurements were taken at 5 K, 30 K, and 150 K using incident energy 12 meV, 50 meV, and 150 meV at each temperature. The powder was sealed with helium exchange gas to ensure good temperature coupling. A powder sample of the nonmagnetic analogue Ba$_3$Lu(BO$_3$)$_3$ was also measured to use for nonmagnetic background subtraction. The neutron data were analyzed using DAVE \cite{DAVE} and the CEF levels were fit using the Python package PyCrystalField \cite{PyCrystalField}.

\begin{figure}
\centering
\includegraphics[width=0.23\textwidth]{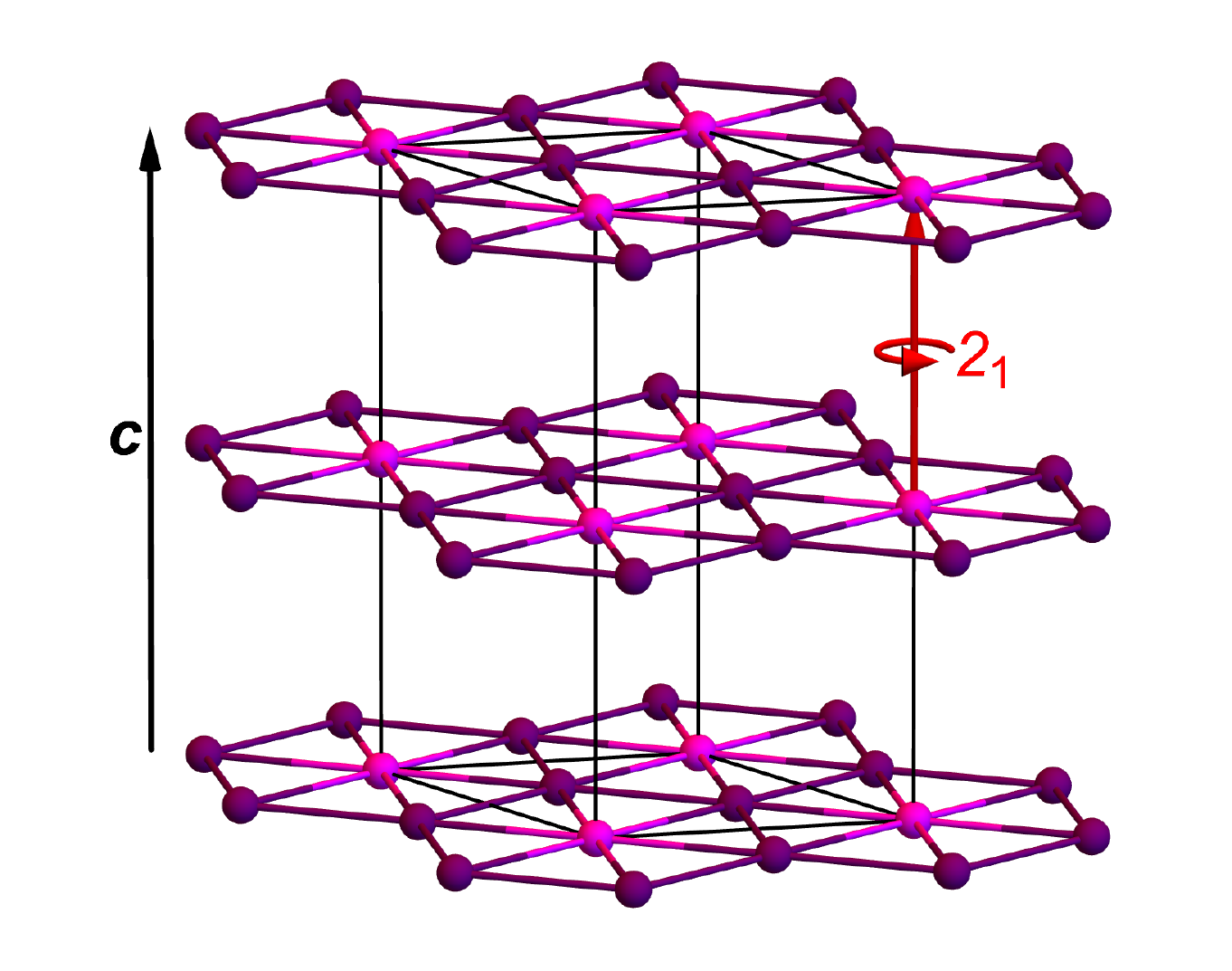}
\includegraphics[width=0.23\textwidth]{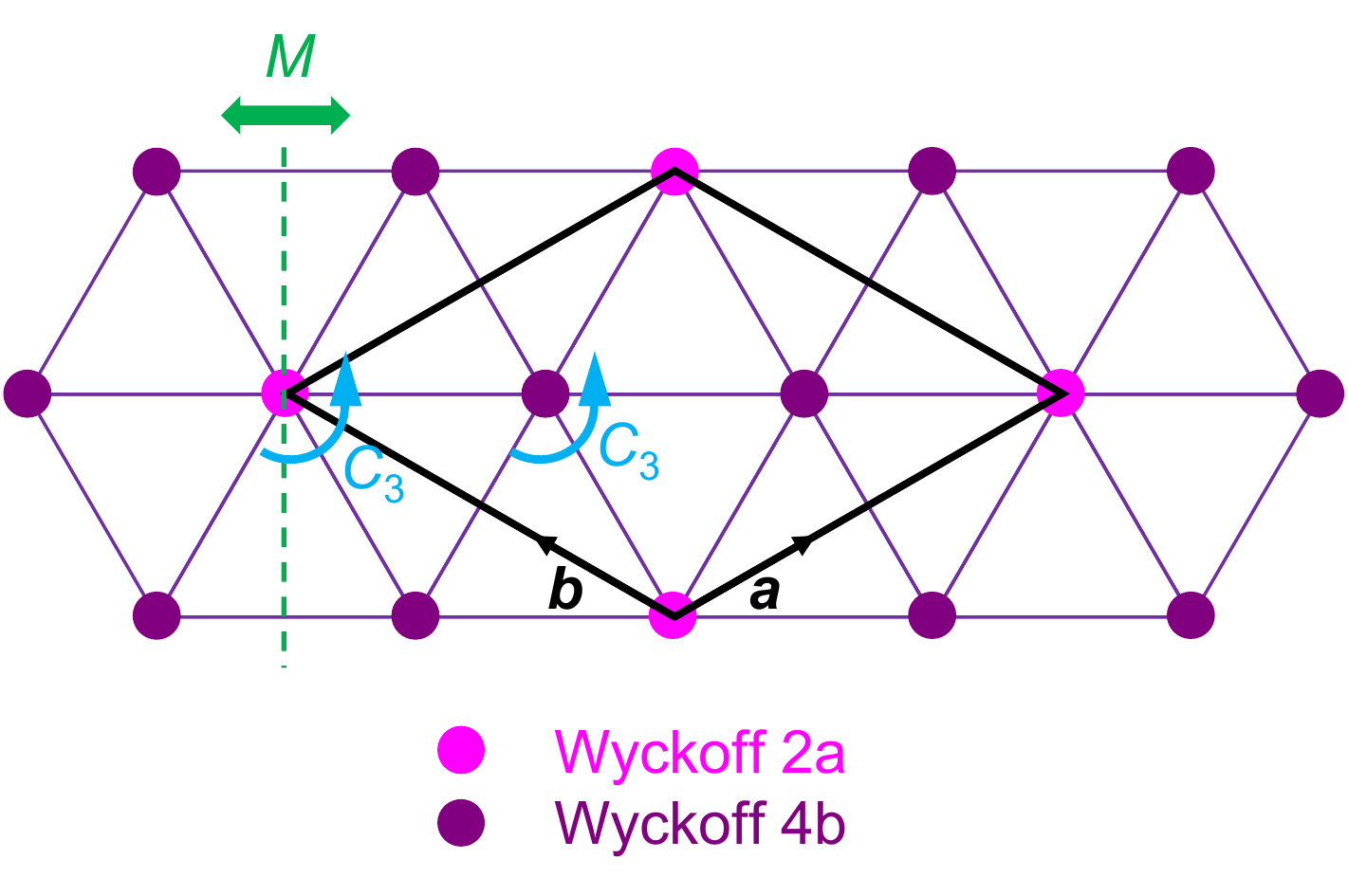}
\caption{Crystal structure, basis vectors and symmetry of Ba$_3$Er(BO$_3$)$_3$. The symmetry operations $2_1$, $C_3$, and $M$ correspond to a 180-degree screw axis in the $z$-direction, a three-fold rotation about the $z$-axis, and a mirror plane, respectively.}\label{fig:BEBO_illus}
\end{figure}

\begin{table}[!thb]
\centering
\caption{Symmetry information for the two environments of Er$^{3+}$.}\label{tab1}
\begin{tabular}{cccc}
\hline
Wyckoff &Coordinates in & Local & Symmetry\\
Position & basis $\bm{a},\bm{b},\bm{c}$& Symmetry  & Generators\\
\hline
\multirow{ 2}{*}{$2a$} & \multirow{ 2}{*}{$(0,0,0)$, $(0,0,\frac{1}{2})$\footnote{More realistic positions are $(0,0,0.00366300)$ and ${1/3, 2/3, 0.49821300}$, see \cite{bebo_lattice_data}.}} & \multirow{ 2}{*}{$\mathrm{C}_{3v}$} & \multirow{ 2}{*}{$C_3$, $M$}\\
&&&\\
\multirow{ 2}{*}{$4b$} & $(\frac{1}{3},\frac{2}{3},0)$, $(\frac{2}{3},\frac{1}{3},0)$, & \multirow{ 2}{*}{$\mathrm{C}_3$} & \multirow{ 2}{*}{$C_3$}\\
&$(\frac{1}{3},\frac{2}{3},\frac{1}{2})$, $(\frac{2}{3},\frac{1}{3},\frac{1}{2})$&&\\
\hline
\end{tabular}
\end{table}

\section{Results and Discussion}
\subsection{Crystal Structure and Symmetry Considerations}
\label{Crystal Structure}
The double borates form two different crystal structures---trigonal with space group R$\Bar{3}$ or hexagonal with space group P6$_3$cm---depending on which rare-earth is used \cite{Ilyukhin1993}. Ba$_3$Er(BO$_3$)$_3$ forms the hexagonal structure with space group P6$_3$cm (space group no.~185) \cite{Khamaganova1999, Cox1994, GaoJAC2018}. The phase purity of Ba$_3$Er(BO$_3$)$_3$ was confirmed using powder x-ray diffraction. Rietveld refinement to the P6$_3$cm structure is shown in the Supporting Information \cite{supplementary}. The corresponding point group is $\mathrm{C}_{6v}$ with 12 elements, generated by a three-fold rotation $C_3$, a two-fold screw $2_1$ along $z$, and a mirror $M$. The lattice structure and symmetries are shown in Fig.~\ref{fig:BEBO_illus}. The unit cell is spanned by the lattice vectors $\bm{a}$, $\bm{b}$, and $\bm{c}$ (see Fig.~\ref{fig:BEBO_illus}) and contains six Er$^{3+}$ ions (two layers $\times$ three Er$^{3+}$ ions in each layer). These Er$^{3+}$ ions occupy two distinct Wyckoff positions and have different symmetry properties, as detailed in Table \ref{tab1}. As we will show, the two environments are important in the understanding of magnetic properties of Ba$_3$Er(BO$_3$)$_3$.

\begin{figure}[htbp]\centering\includegraphics[width= 8.5 cm]{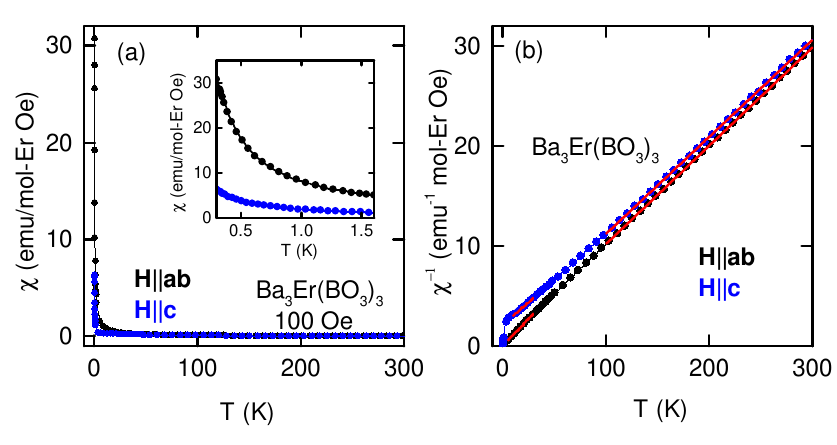}
	\caption{(a) Magnetic susceptibility data for a single crystal sample with field applied in-plane ($\chi_{ab}$) and along the $c$-axis ($\chi_c$). The inset shows the low-temperature region from 300 mK to 1.6 K, showing the anisotropy in the system. (b) Inverse susceptibility data. Red lines show the fit to the Curie--Weiss law.}
	\label{Susceptibility}
\end{figure}

\subsection{Magnetic Susceptibility and Magnetization}
Magnetic susceptibility measurements were performed on a single crystal sample with field applied both parallel to the $c$-axis and parallel to the $ab$-plane down to 300 mK using helium-3 insert, as shown in Fig. \ref{Susceptibility}(a). As shown in the inset of Fig. \ref{Susceptibility}(a), in both orientations we observe no sharp features in the susceptibility down to 300 mK that would indicate magnetic ordering. The inverse susceptibility data have been fit to the Curie--Weiss law: $\chi^{-1} = \frac{T-\theta}{C}$, where $\theta$ is the Curie--Weiss temperature and $C$ is the Curie constant. The inverse susceptibility data and Curie--Weiss fits are shown in Fig. \ref{Susceptibility}(b). We performed fits in the high-temperature region (100 K - 300 K) and in the low-temperature region (10 K - 30 K). The fit parameters were used to compute the effective magnetic moment ($\mu_{eff}$) of the Er$^{3+}$ ions. The Curie--Weiss fit in the high-temperature region was also used to calculate the Land\'{e} $g$-factor. 

In Ba$_3$Er(BO$_3$)$_3$ we observe strong directional anisotropy in the susceptibility data. We do not observe any discrepancy between the field-cooled and zero field-cooled data in either orientation. In previously reported triangular compounds Ba$_3$Yb(BO$_3$)$_3$ and NaYbBa(BO$_3$)$_2$, large anisotropy is also observed \cite{BYBOPRB, Guo_2019}. In these compounds the intralayer Yb--Yb distance is less the interlayer distance, and the susceptibility and magnetization is larger with field applied along the c-axis. Similarly in Ba$_3$Er(BO$_3$)$_3$, the $\chi_c$ value is expected to be higher than $\chi_{ab}$ since the intralayer Er--Er distance is smaller than the interlayer Er--Er distance (see Table \ref{tab2}). This is observed for NaErBa(BO$_3$)$_2$ \cite{Guo_MRE2019}, however for Ba$_3$Er(BO$_3$)$_3$ the trend is reversed: the susceptibility $\chi_c$ is smaller than $\chi_{ab}$, particularly at low temperatures, as shown in Fig. \ref{Susceptibility}(a). A similar anisotropy ($\chi_{ab} > \chi_c$) is reported for other Er based systems AErSe$_2$ (A = Na, K) \cite{XingPRM2019}.  As shown in Fig. \ref{Susceptibility}(b), the inverse susceptibility curves for both orientations are nearly parallel from 100 K to 300 K, with $\theta_{ab}$ = -3.75 K and $\theta_{c}$ = -15.46 K. These values are notably smaller than those observed in ErMgGaO$_4$ \cite{Cevallos_SSC2018}, but are similar to the values observed in the Er chalcogenides NaErCh$_2$ (Ch = O, S, Se) \cite{Hashimoto_2003, LiuChinLett2018, XingPRM2019}. The calculated effective moments and Land\'e $g$-factors in this region are $\mu_{eff}(ab)$ = 9.02 $\mu_B$ and $g_{ab}$ = 1.13 for $H\parallel ab$ and $\mu_{eff}(c)$ = 9.07 $\mu_B$ and $g_{c}$ = 1.13 for $H\parallel c$ , in good agreement with the $\mu = 9.58 \mu_B$ and $g = 1.2$ expected for a free Er$^{3+}$ ion \cite{BlundellMagnetism}. 

\begin{figure}[htbp]\centering\includegraphics[width= 8.5 cm]{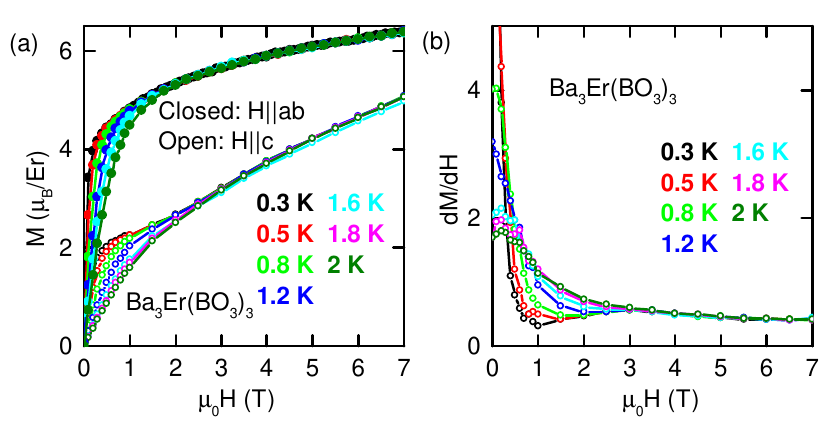}
	\caption{(a) Isothermal magnetization data of Ba$_3$Er(BO$_3$)$_3$ taken at various temperatures with field applied in-plane and along the $c$-axis. (b) First derivative of isothermal magnetization with field applied along the $c$-axis showing anomaly between 1 T and 3 T.}
	\label{Magnetization}
\end{figure}

With field applied in the $ab$-plane, the inverse susceptibility remains nearly linear down all the way to 300 mK. In contrast, the perpendicular direction is linear down to around 10 K, where it levels off slightly before turning down sharply at 5 K. This downturn behavior is also observed in Yb-based triangular lattice compounds, such as Ba$_3$Yb(BO$_3$)$_3$, NaYbO$_2$, and NaBaYb(BO$_3$)$_3$ \cite{BYBOPRB, GuoPRM2020, LiuChinLett2018, Guo_2019}, and is believed to be caused by thermal population of CEF levels. The Curie--Weiss fits from 10 K to 30 K give $\theta_{ab} = -0.1876$ K and $\mu_{eff}(ab) = 8.51 \mu_B$ for $H\parallel ab$, showing a slight decrease in both values for both parameters, and $\theta_{c} = -24.16$ K and $\mu_{eff}(c) = 9.50 \mu_B$ for $H\parallel c$. Below 5 K the parameters sharply decrease. The inverse susceptibility remaining nearly linear down to very low temperatures is observed in several other Er-based triangular lattice systems \cite{Sanders_2017, Cevallos_SSC2018, Hashimoto_2003}, and the discrepancy between the low temperature behavior for the two orientations is also seen in single crystals of NaErSe$_2$ \cite{XingPRM2019}.

To further investigate the magnetic behavior of this material, isothermal magnetization measurements were carried out at different temperatures ranging from 300 mK to 2 K with field up to 7 T applied parallel to the $c$-axis and parallel to the $ab$-plane, as shown in Fig. \ref{Magnetization}(a). For both orientations there is a nonlinear region at low field where the magnetization rapidly increases before the curve becomes nearly linear above 3 T. With field applied in the $ab$-plane, the magnetization increases quickly at first and then slowly tapers off as the field increases. In contrast, with field applied along the $c$-axis the magnetization begins to level off around 1 T before increasing again. This behavior can be seen more clearly in the plot of $dM/dH$ shown in Fig. \ref{Magnetization}(b). The $dM/dH$ curve goes down to a local minimum at 1 T, then rises to a local maximum at 3 T before going back down at higher fields. As with susceptibility, the magnetization data shows strong anisotropy in this material. Similar anisotropy in the saturated magnetization and curve shape are observed in ErMgGaO$_4$ (where a crossing of the magnetization curves is also observed) and AErSe$_2$ (A = Na and K) \cite{Cevallos_SSC2018, XingPRM2019}.

\begin{figure}
    \centering
    \includegraphics[width=8.5 cm]{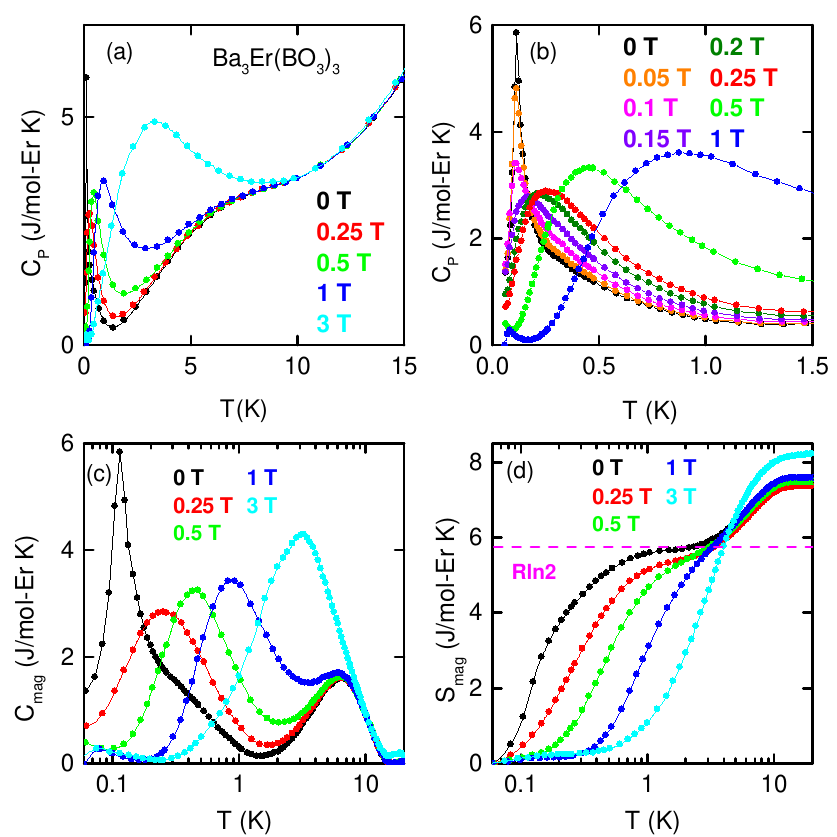}
    \caption{(a) Heat capacity of Ba$_3$Er(BO$_3$)$_3$ at various fields applied along $c$-axis, up to 15 K. (b) Low-temperature heat capacity data with field applied in smaller increments. (c) Magnetic heat capacity of Ba$_3$Er(BO$_3$)$_3$ obtained by subtracting $T^3$ fit to the 0 T data. (d) Magnetic entropy up to 20 K.}
    \label{HC}
\end{figure}

\subsection{Heat Capacity}
To further characterize this material, we also performed heat capacity measurements at various fields from room temperature down to 60 mK using helium-4 and dilution refrigerator (DR) setups. The measurements were carried out on a single crystal sample of Ba$_3$Er(BO$_3$)$_3$ with field applied parallel to the $c$-axis. The heat capacity data up to 15 K with several applied fields (0 T, 0.25 T, 0.5 T, 1 T, 3 T) are shown in Fig. \ref{HC}(a). The zero field data shows a sharp peak at 0.1 K, possibly indicating the onset of magnetic order. The peak broadens and shifts to higher temperature with increasing field strength, and appears more like a two-level Schottky anomaly. In order to further study the transition between the sharp peak and broad features, we measured the heat capacity while increasing the applied field in small steps. For clarity, the fields below 1 T are shown separately in Fig. \ref{HC}(b) for the low-temperature region up to 1.5 K. As the field is slowly increased up to 0.1 T, the intensity of the peak decreases but the peak's position at 0.1 K does not change. As the field is further increased beyond 0.15 T, the peak begins to broaden and shift to higher temperatures, matching the behavior we observed for the higher fields. The magnetic heat capacity and corresponding magnetic entropy are shown in Fig. \ref{HC}(c) and (d). The lattice contribution to the heat capacity was estimated by fitting the zero field data to a cubic model ($T^3$), which was subtracted to obtain the magnetic heat capacity ($C_{mag}$). The magnetic heat capacity of Ba$_3$Er(BO$_3$)$_3$ reveals a low-temperature peak that shifts to higher temperature with increasing field, which lines up with the Schottky-like feature we observe. We also observe a peak at 6.5 K which is present for all fields measured.

The entropy as a function of temperature [see Fig.~\ref{HC} (d)] shows a plateau at around 2 K, with a value of $R \ln 2\sim 5.76$ $  \text{J}/\text{mol}\cdot \text{K}$. This indicates that there is an effective two-level system at very low energy, agreeing with the prediction of a Kramers' doublet ground state \cite{Scheie_2020,Gao_2020}. Furthermore, another plateau is observed at about 10 K, with an entropy value about $\frac{4}{3} R \ln 2 \sim 7.68 \text{J}/\text{mol}\cdot \text{K}$. If we assume a CEF doublet exists at 6.5 K where the magnetic heat capacity $C_{\text{mag}}$ has a peak, then this CEF doublet is associated to only one-third of the Er ions \--- this is the only possibility in order to obtain a $\frac{4}{3} R \ln 2$ entropy. We associate this CEF excitation \emph{with the Er atoms at Wyckoff position $2a$}, which indeed occupy one-third of the total Er atoms.

\

\subsection{Neutron Scattering Measurements}
\begin{figure}
    \centering
    \includegraphics[width=0.43\textwidth]{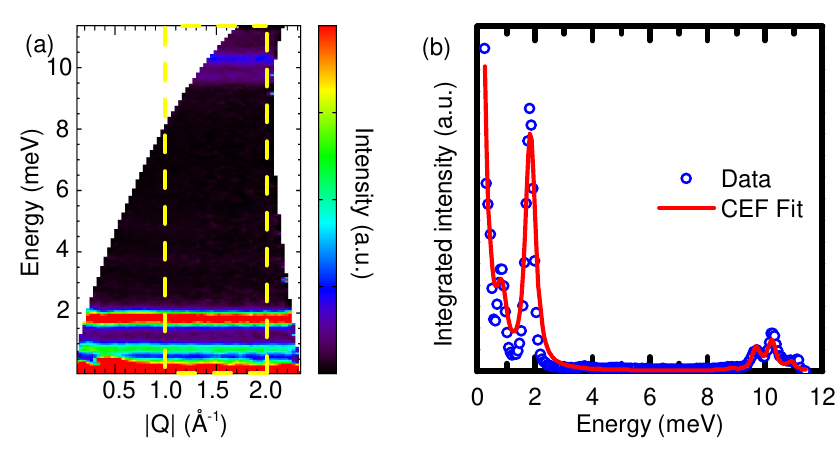}
    \caption{(a) Inelastic neutron scattering data taken on a powder sample of Ba$_3$Er(BO$_3$)$_3$ at 5 K. (b) Integrated intensity from the boxed region in (a), along with Stevens operator fit to the data.}
    \label{CEF}
\end{figure}

To get a better understanding of the CEF spectrum of Ba$_3$Er(BO$_3$)$_3$, inelastic neutron scattering measurements were carried out on a powder sample at the SEQUOIA time-of-flight spectrometer at Oak Ridge National Laboratory \cite{SEQUOIA}. Measurements were performed with incident energies 150 meV, 50 meV, and 12 meV, at temperatures 5 K, 30 K and 150 K. Er${}^{3+}$ has $S = 3/2$---so all of its energy levels are Kramers doublets---and has $J = 15/2$---so one would expect to see at total of 7 excitations above the ground state. Our data shows two low-energy excitations at 0.8 and 1.9 meV and two higher-energy excitations around 10 meV as shown in Fig. \ref{CEF}(a). Additionally, we see signatures of two closely lying excitations at higher energies of approximately 12.5 and 13.5 meV -- see Fig. S2. The neutron spectrum with $E_i = 12 \text{meV}$ taken at 5 K and a line cut to show the intensity as a function of energy are shown in Fig. \ref{CEF}(b).  The first excitation at 0.8 meV corresponds to a temperature of 9.3 K which coincides with the broad feature we observe in the magnetic heat capacity measurements [see Fig. \ref{HC}(c)]. The next excitation occurs at 1.9 meV which corresponds to a temperature of 22 K, which matches with the small peak in the magnetic heat capacity we see at the highest temperatures of our measurement. The CEF spectrum we observe is similar to the values calculated for ErMgGaO$_4$ and Ba$_3$ErB$_9$O$_{18}$, (although the excitations tend to appear at slightly lower energies in Ba$_3$Er(BO$_3$)$_3$ than in either of these materials) \cite{Cai_muSR2020, Khatua_PRB2022}, and for those observed in NaErS$_2$ and KErSe$_2$ \cite{Gao_2020, Scheie_2020}.

The low energy levels observed in the magnetic heat capacity and inelastic neutron scattering are captured by the CEF Hamiltonian $H$. The two Er sites have different point symmetries, so depending on the local symmetry being $\mathrm{C}_3$ (for 4b positions) or $\mathrm{C}_{3v}$ (for 2a positions), we have
\begin{widetext}
\begin{subequations}
\begin{align}
H_{4b} &= B^0_2 O_{2,0} + B^0_4 O_{4,0} + B^3_4 O_{4,3} + B^{-3}_4 O_{4,-3}
+
B^0_6 O_{6,0} + B^3_6 O_{6,3} + B^{-3}_6 O_{6,-3} + B^6_6 O_{6,6} + B^{-6}_6 O_{6,-6},
\label{eq:H2a}\\
H_{2a} &= B^0_2 O_{2,0} + B^0_4 O_{4,0} + B^3_4 O_{4,3} + B^0_6 O_{6,0} + B^3_6 O_{6,3}  + B^6_6 O_{6,6},
\end{align}
\end{subequations}
\end{widetext}

Here $O_{n,m}$ are the standard Stevens operators \cite{Stevens_1952}, and $B^m_n$ are the Stevens parameters to be determined. We perform a point charge calculation to obtain the Stevens parameters and the CEF levels \cite{Hutchings1964}. See Ref.~\cite{bauer2010magnetism} for derivation. For the ion positions we use the data from \cite{bebo_lattice_data}. We consider the following two models:
\begin{itemize}
\item The Ba-Er-B-O model: We calculate the CEF contribution from all types of ions. 
\item The O-only model: We only include the CEF contribution from the oxygen ions O$^{2-}$. This describes the scenario in which the CEF  is mainly due to the ligands. 
\end{itemize}

\begin{table*}
\caption{CEF data obtained from the point-charge models and comparison to the experimental data.}\label{tab:bebo_cef}
\begin{tabular}{cccccccccc}
\hline
\multicolumn{5}{l}{\textbf{Model:  Ba-Er-B-O}}\\
&&\multicolumn{8}{l}{2a levels  (meV): $E_1=4.390$, $E_2=12.43$, $E_3=22.79$, $E_4=34.54$, $E_5=49.44$, $E_6=55.93$, $E_7=57.67$}\\
&&\multicolumn{8}{l}{4b levels (meV): $E_1 = 4.458$, $E_2=12.59$, $E_3=22.95$, $E_4=34.63$, $E_5=49.51$, $E_6=55.71$, $E_7=57.41$}\\
\hline
Position& $B^0_2$ & $B^0_4$ & $B^3_4$ & $B^{-3}_4$ & $B^0_6$ & $B^3_6$ & $B^{-3}_6$ & $B^6_6$ & $B^{-6}_6$\\
\hline
2a & $0.293$ & $ -0.000572$&$ 0.0146$&$ 0$&$ 8.30\times 10^{-7}$ &$ 0.0000320$&$ 0$&$ 9.91\times 10^{-6} $&$ 0$\\
4b & $0.288$&$ -0.000578$&$ 0.0149$&$ -0.00105$&$ 9.26\times 10^{-7}$&$ 0.0000313$&$-2.11\times 10^{-6}$&$ 0.0000103$&$ -1.45\times 10^{-6}$\\
\hline
\hline
\multicolumn{5}{l}{\textbf{Model:  O-only}}\\
&&\multicolumn{8}{l}{2a levels (meV): $E_1=0.2322$, $E_2=6.977$, $E_3=7.440$, $E_4=9.200$, $E_5=32.44$, $E_6=32.53$, $E_7=33.10$}\\
&&\multicolumn{8}{l}{4b levels (meV):  $E_1=1.156$, $E_2=7.748$, $E_3=7.973$, $E_4=9.796$, $E_5=33.35$, $E_6=33.50$, $E_7=33.94$}\\
\hline
Position& $B^0_2$ & $B^0_4$ & $B^3_4$ & $B^{-3}_4$ & $B^0_6$ & $B^3_6$ & $B^{-3}_6$ & $B^6_6$ & $B^{-6}_6$\\
\hline
2a &$ 0.0190$&$ -0.000688$&$ 0.0187$&$ 0$&$ 1.23\times 10^{-6}$&$ 0.0000350$&$ 0$&$ 0.0000150$&$ 0$\\
4b & $0.00511$&$ -0.000696$&$0.0190$&$ -0.00126$&$ 1.34\times 10^{-6}$&$ 0.0000343$&$-2.36\times 10^{-6}$&$ 0.0000154$&$ -2.03\times 10^{-6}$\\
\hline
\hline
\multicolumn{5}{l}{\textbf{Model:  O-only, with radial factor $\gamma=1.1$}}\\
&&\multicolumn{8}{l}{2a levels (meV): $E_1=0.7997$, $E_2=9.299$, $E_3=9.832$, $E_4=11.39$, $E_5=42.19$, $E_6=42.70$, $E_7=43.42$}\\
&&\multicolumn{8}{l}{4b levels (meV):  $E_1=1.818$, $E_2=10.22$, $E_3=10.53$, $E_4=12.08$, $E_5=43.31$, $E_6=43.90$, $E_7=44.49$}\\
\hline
Position& $B^0_2$ & $B^0_4$ & $B^3_4$ & $B^{-3}_4$ & $B^0_6$ & $B^3_6$ & $B^{-3}_6$ & $B^6_6$ & $B^{-6}_6$\\
\hline
2a &$ 0.0215$&$ -0.000893$&$ 0.0243$&$ 0$&$ 1.80\times 10^{-6}$&$ 0.0000512$&$ 0$&$ 0.0000220$&$ 0$\\
4b & $0.00579$&$ -0.000904$&$0.0246$&$ -0.00163$&$ 1.96\times 10^{-6}$&$ 0.0000502$&$-3.46\times 10^{-6}$&$ 0.0000226$&$ -2.96\times 10^{-6}$\\
\hline
\hline
\multicolumn{5}{l}{\textbf{Direct fit to neutron data}}\\
&&\multicolumn{8}{l}{2a levels (meV): $E_1=0.850$, $E_2=9.70$, $E_3=10.91$, $E_4=12.55$, $E_5=45.11$, $E_6=45.48$, $E_7=46.34$}\\
&&\multicolumn{8}{l}{4b levels (meV):  $E_1=1.850$, $E_2=10.25$, $E_3=12.50$, $E_4=13.32$, $E_5=48.22$, $E_6=49.65$, $E_7=50.39$}\\
\hline
Position& $B^0_2$ & $B^0_4$ & $B^3_4$ & $B^{-3}_4$ & $B^0_6$ & $B^3_6$ & $B^{-3}_6$ & $B^6_6$ & $B^{-6}_6$\\
\hline
2a &$ 0.0186$&$ -0.000910$&$ 0.02639$&$ 0$&$ 1.77\times 10^{-6}$&$ 0.0000510$&$ 0$&$ 0.0000242$&$ 0$\\
4b & $0.00759$&$ -0.001039$&$-0.002126$&$ -0.02638$&$ 2.402\times 10^{-6}$&$ -4.908\times 10^{-6}$&$-0.0000684$&$ -2.772\times 10^{-6}$&$ -3.931\times 10^{-7}$\\
\hline
\end{tabular}
\end{table*}

The results are shown in Table \ref{tab:bebo_cef}. The Ba-Er-B-O model gives comparable CEF levels for Er$^{3+}$ in both the 2a and 4b environments. On the other hand, the O-only model has the first CEF excitation for the 2a position at $E_1 = 0.2322$ meV (2.7 K) and for the 4b position at 1.156 meV (13.4 K); as a result, it provides one scenario where the 2a has a significantly lower first CEF excitation, in agreement with the CEF result from the entropy analysis. We point out that the point charge model so far involves no fitting parameters, and the qualitative agreement with the experiment is remarkable.

To further quantitatively improve our result, we include a radial factor $\gamma$ in the O-only model to control the ion's radius. As one can see, the O-only model with a radial factor $\gamma=1.1$ reproduces the low energy CEF levels extracted experimentally quite well. For comparison, we also fit CEF Hamiltonian directly to our neutron data, using the point charge calculations as a guide for which peaks to fit to which ion.  The result of that fit is also given in Table \ref{tab:bebo_cef} for a comparison and the fit is shown in Fig. \ref{CEF}(b). However, we note that this fit is likely unreliable: since we only observe 4 distinct peaks in our data, the fit is largely unconstrained and gives several additional transitions at higher energies which we do not observe in our neutron data. The full set of neutron spectra and comparisons of the 3 models to our data are presented in the Supplemental Materials \cite{supplementary}. While the O-only model seems to give a good estimate of the CEF levels, we point out two potential issues: (1) it is yet to find a reason why keeping only the anions (O$^{2-}$) improves the modeling. Naively this describes the situation where the positive charges from the cations are maximally screened; such screening effect is beyond a point charge description. (2) We note that the Stevens parameters from the point charge model show a large difference from the experimental values (particularly in the values of the 4a Er site). More experimental data are needed in order to better understand the crystal field physics. To summarize, a point charge model for the CEF levels from only O$^{2-}$ contributions qualitatively reproduces the magnetic heat capacity at low temperature due to the ground state Kramers pair and low energy first excitation Kramers pair only for Wyckoff position 2a.

\subsection{Exchange physics}

The low energy physics in the temperature range 0 K$<T<$ 11 K happens in the ground state doublet and the first excited doublet for the 2a position ions. If we restrict to a smaller temperature range 0 K$<T<$ 2 K, the physics happens purely in the ground state doublet. In this section we focus on the smaller temperature range 0 K$<T<$ 2 K. Our goal is to analyze the possible spin Hamiltonian consistent with the experimentally observed magnetic heat capacity. At nearest-neighbor level there are two types of exchanges: those connecting two 4b sites, which we call $J_{4b}$, and those connecting 2a and 4b, which we call $J_{2a}$. See Table \ref{tab2} for more information. The triangular lattice can be viewed as a honeycomb lattice of 4b ions decorated by a (larger) triangular lattice of 2a ions located at the hexagon centers; see Fig.~\ref{fig:BEBO_illus} for an illustration. The exchanges $J_{4b}$ and $J_{2a}$ can have different energy scales. 

\subsubsection{2D Magnetic order}

Two pieces of information can be extracted from the magnetic heat capacity data in Fig.~\ref{HC}(b).  First, the peak at $\sim 0.1$ K does not immediately vanish at small magnetic field, resembling an \emph{antiferromagnetic ordering transition}.    Second, there is a lower, broader peak at $\sim 0.3$ K, evident at small magnetic field, that moves towards higher temperatures as the field increases. We associate this broader peak with \emph{ferromagnetic fluctuations} of the system. Based on this analysis, we propose that {Ba$_3$Er(BO$_3$)$_3$ realizes interesting two-sublattice exchange physics, in which the honeycomb lattice spins develop ferromagnetic correlations due to the additional spins at the hexagon centers but eventually order in an antiferromagnetic phase.}

\subsubsection{3D magnetic order}

Even if the  magnetic order is determined within each layer,  nontrivial ordering can still happen between the layers. To study the possible 3D magnetic order we now determine the magnetic symmetry that is consistent with the 2D magnetic order. Within each layer, the magnetic order has three-fold rotation symmetry $C_3$ and the magnetic mirror symmetry $M_x'=M\mathcal{T}$ (product of the ordinary mirror reflection $M$ given in \ref{tab1} and time reversal $\mathcal{T}$). Furthermore, the two layers of Er$^{3+}$ in one unit cell can still have different order configurations: the order on the two layers can be either related by $2_1$ (the ordinary screw in Table \ref{tab1}) or $2'_1 = 2_1\cdot\mathcal{T}$. Physically, these two cases correspond to antiferromagnetic and ferromagnetic (collinear) order between the layers. The corresponding magnetic space groups are P6$_3'$cm and P6$_3$c$'$m$'$ \cite{litvin2008tables}, respectively. Note that both cases allow an effective Zeeman field on the $2a$ spin:
\begin{equation}
H_{\text{Zeeman}} = h^z \sum_i  S^z_{i,2a},
\end{equation}
where $h^z$ is the effective Zeeman field allowed by the magnetic space group symmetry at the $2a$ site. The isothermal magnetization measurement (Fig.~\ref{Magnetization}) observes zero magnetization at zero field for temperatures all the way down to 0.3 K (the lowest temperature reached). One scenario consistent with this is that the two layers order antiferromagnetically, corresponding to the magnetic space group P6$_3'$cm.

\begin{table}
\centering
\caption{Distances between Er$^{3+}$ ions and definition of exchange parameters.}\label{tab2}
\begin{tabular}{c|ccc}
\hline
Intra-layer bonds  & 4b-4b & 4b-2a & 2a-2a and 4b-4b \\
   Bond distance       & 5.47315 \AA & 5.47403 \AA & 9.00139 \AA\\
Exchange energy & $J_{4b}$ & $J_{2a}$ & $J_3$ \\
          \hline
Inter-layer bond & 4b-4b and 2a-2a\\
Bond distance   & 9.48007 \AA \\
Exchange energy & $J_{u-d}$  \\
          \hline
\end{tabular}
\end{table}

\subsubsection{Exchange physics}

In this section we discuss the influence of exchange interactions in the context of the experimental data.  Based on the structure and symmetry, a minimal exchange model consists of the Ising hamiltonian  \footnote{We note that a related model with Heisenberg exchange interaction \cite{PhysRevB.104.134432} may also be relevant to our compound Ba$_3$Er(BO$_3$)$_3$.}
\begin{equation}\label{inlayerIsing}
H_{\text{Ising}}
=
J_{2a} \sum_{i \in 2a,j\in 4b}  S^z_i S^z_j
+J_{4b} \sum_{i,j \in 4b} S^z_iS^z_j 
+H^z \sum_iS^z_i,
\end{equation}
with antiferromagnetic interactions $J_{2a}>0$, $J_{4b}>0$. This model contains several simpler limits.  First, for $J_{2a}=J_{4b}$ it reduces to the triangular lattice Ising model, which is a canonical problem in statistical physics, famous for its frustration and lack of magnetic order.  Second, if we include only $J_{4b}>0$ exchange, it reduces to a honeycomb lattice Ising model on the 4b sites, which is unfrustrated and shows an antiferromagnetic ordering transition. 
The phase diagram of the model \eqref{inlayerIsing} at zero-field ($H^z=0$) can be obtained in closed form and contains three phases \cite{PhysRevB.43.8759}. At nonzero field $H^z>0$, the phase diagrams of the two limiting cases above both contain an ordered phase and a disordered phase \cite{PhysRevLett.62.2773,METCALF19731}.

We now discuss the experimental heat capacity shown in Fig.~\ref{HC} in context of exchange physics. Ideally, one would like to explain the following key features of the data: At low field there is a singular peak at about $100$ mK presumably reflecting long range ordering.  This peak decreases in magnitude with weak fields, and becomes noticeably broaden above $0.1$ T.  Above this field, the now broadened low-temperature peak moves upward in temperature with increasing field.  In addition, a second distinct broad high temperature (at around $6.5$ K in zero field) peak is present at all fields up to about $2$ T, when the peak originating at low temperature merges with it.  

The persistent sharpness of the low-$T$ peak in non-zero fields below $0.1$ T, shown in Fig.~\ref{HC}(b), suggests that it is antiferromagnetic in nature, distinct from the peak in an Ising ferromagnet which broadens immediately upon application of a field.  By contrast, in an antiferromagnet, the ordering peak typically weakens and initially moves to lower temperature upon application of a field, due to the competition of antiferromagnetism, which wishes to anti-align the moments and the field, which tries to align them.  At somewhat larger fields, the broadened peak typically reverses direction and increases its temperature, as the Zeeman term increases the thermal energy required to release the magnetic entropy. These behaviors are common to many antiferromagnetic models: The low field suppression of ordering temperature happens in the honeycomb lattice Ising model \cite{KIM2006245}, and the high field enhancement of the heat capacity peak and peak temperature happens in both the honeycomb lattice and the triangular lattice Ising models \cite{doi:10.1143/JPSJ.53.3060}.

In the experiment, we indeed observe an initial weaking of the ordering peak, but not a shift to lower temperatures.  This is perhaps due to the very low temperature of the measured ordering transition.  A broadened low-temperature peak moving to higher temperatures for $H>0.15$ T is consistent with the expectations just described. As discussed earlier, the high-temperature $6.5$ K peak may originate from the crystal field excitations of the 2a sites.

Given the rough consistency of the picture of weak antiferromagnetism, we briefly consider the nature of the possible ordered state.   We further introduce an Ising exchange coupling between the two layers, and the total Hamiltonian is
\begin{equation}\label{twolayerIsing}
H_{\text{total}} = H_{\text{Ising},u} + H_{\text{Ising},d} + J_{u-d}\sum_i S^z_i S^z_j.
\end{equation}
In general this model results in three-dimensional ordering.  In the case of zero field, its phase diagram is a simple extension of the 2D phase diagram \cite{PhysRevB.43.8759}. We conclude that for $J_{4b}>J_{2a}$, $J_{u-d}>0$, it supports the following candidate three-dimensional magnetic structure below the zero field transition: Neel order on the honeycomb lattice decorated with extra spins aligned ferromagnetically on the hexagon centers, while the layers order antiferromagnetically with a two-layer periodicity. Further magnetic measurements below the $100$ mK ordering transition will be required to test and refine this proposed structure.

\section{Conclusions}
\label{Conclusions}
We have synthesized high-quality single crystal samples of the erbium-based triangular lattice compound Ba$_3$Er(BO$_3$)$_3$. Low temperature magnetic measurements reveal the presence of large anisotropy in this system. The zero field magnetic entropy as a function of temperature shows two plateaus, one ($T= 2$ K) at $R \ln 2$ and the other ($T= 10$ K) at $\frac{4}{3} R \ln 2$. We show that these can be understood from the two environments that the magnetic ions Er$^{3+}$ occupy (Wyckoff positions 2a and 4b) with distinct symmetry properties and crystal field environments. A point charge calculation for the crystal electric field levels is consistent with this understanding. Furthermore, we discuss the exchange physics happening in the ground state doublet.   The material exhibits a zero field ordering transition, which we argue to be antiferromagnetic in nature, and several features of the heat capacity can be understood in this context.  

\

\section{Acknowledgements}
Research performed at Duke University is supported by National Science Foundation Grant No. DMR-1828348. A portion of this research used resources at the Spallation Neutron Source, a DOE Office of Science User Facility operated by the Oak Ridge National Laboratory. R.B. acknowledges the support provided by Fritz London Endowed Post-doctoral Research Fellowship at Duke University. S.H. acknowledges the support provided by William M. Fairbank chair in Physics at Duke University. C.L. acknowledges the fellowship support of the Gordon and Betty Moore Foundation through the Emergent Phenomena in Quantum Systems (EPiQS) program.  L.B. was supported by the DOE, Office of Science, Basic Energy Sciences under Award No. DE-FG02-08ER46524, and by the Simons Collaboration on Ultra-Quantum Matter, which is a grant from the Simons Foundation (651440).

\section{Author Contributions}
Research conceived by S.H.; M.E., R.B., S.E.D. and S.H. synthesized samples; M.E., R.B. and S.H. performed thermodynamics measurements; M.E., S.E.D., A.I.K and S.H. conducted neutron scattering experiments; C.L. and L.B. provided theoretical interpretations; M.E., R.B., C.L., L.B. and S.H. wrote the manuscript with comments from all authors.

\bibliographystyle{apsrev4-2}
\bibliography{BErBO}

\begin{thebibliography}{62}%
\makeatletter
\providecommand \@ifxundefined [1]{%
 \@ifx{#1\undefined}
}%
\providecommand \@ifnum [1]{%
 \ifnum #1\expandafter \@firstoftwo
 \else \expandafter \@secondoftwo
 \fi
}%
\providecommand \@ifx [1]{%
 \ifx #1\expandafter \@firstoftwo
 \else \expandafter \@secondoftwo
 \fi
}%
\providecommand \natexlab [1]{#1}%
\providecommand \enquote  [1]{``#1''}%
\providecommand \bibnamefont  [1]{#1}%
\providecommand \bibfnamefont [1]{#1}%
\providecommand \citenamefont [1]{#1}%
\providecommand \href@noop [0]{\@secondoftwo}%
\providecommand \href [0]{\begingroup \@sanitize@url \@href}%
\providecommand \@href[1]{\@@startlink{#1}\@@href}%
\providecommand \@@href[1]{\endgroup#1\@@endlink}%
\providecommand \@sanitize@url [0]{\catcode `\\12\catcode `\$12\catcode
  `\&12\catcode `\#12\catcode `\^12\catcode `\_12\catcode `\%12\relax}%
\providecommand \@@startlink[1]{}%
\providecommand \@@endlink[0]{}%
\providecommand \url  [0]{\begingroup\@sanitize@url \@url }%
\providecommand \@url [1]{\endgroup\@href {#1}{\urlprefix }}%
\providecommand \urlprefix  [0]{URL }%
\providecommand \Eprint [0]{\href }%
\providecommand \doibase [0]{https://doi.org/}%
\providecommand \selectlanguage [0]{\@gobble}%
\providecommand \bibinfo  [0]{\@secondoftwo}%
\providecommand \bibfield  [0]{\@secondoftwo}%
\providecommand \translation [1]{[#1]}%
\providecommand \BibitemOpen [0]{}%
\providecommand \bibitemStop [0]{}%
\providecommand \bibitemNoStop [0]{.\EOS\space}%
\providecommand \EOS [0]{\spacefactor3000\relax}%
\providecommand \BibitemShut  [1]{\csname bibitem#1\endcsname}%
\let\auto@bib@innerbib\@empty
\bibitem [{\citenamefont {Anderson}(1973)}]{AndersonMatRes1973}%
  \BibitemOpen
  \bibfield  {author} {\bibinfo {author} {\bibfnamefont {P.}~\bibnamefont
  {Anderson}},\ }\href
  {https://doi.org/https://doi.org/10.1016/0025-5408(73)90167-0} {\bibfield
  {journal} {\bibinfo  {journal} {Materials Research Bulletin}\ }\textbf
  {\bibinfo {volume} {8}},\ \bibinfo {pages} {153 } (\bibinfo {year}
  {1973})}\BibitemShut {NoStop}%
\bibitem [{\citenamefont {Balents}(2010)}]{BalentsNature2010}%
  \BibitemOpen
  \bibfield  {author} {\bibinfo {author} {\bibfnamefont {L.}~\bibnamefont
  {Balents}},\ }\href {https://doi.org/10.1038/nature08917} {\bibfield
  {journal} {\bibinfo  {journal} {Nature}\ }\textbf {\bibinfo {volume} {464}},\
  \bibinfo {pages} {199} (\bibinfo {year} {2010})}\BibitemShut {NoStop}%
\bibitem [{\citenamefont {Li}\ \emph {et~al.}(2015{\natexlab{a}})\citenamefont
  {Li}, \citenamefont {Liao}, \citenamefont {Zhang}, \citenamefont {Li},
  \citenamefont {Ling}, \citenamefont {Zhang}, \citenamefont {Zou},
  \citenamefont {Yang}, \citenamefont {Wang},\ and\ \citenamefont
  {Wu}}]{LiSciR2015}%
  \BibitemOpen
  \bibfield  {author} {\bibinfo {author} {\bibfnamefont {Y.}~\bibnamefont
  {Li}}, \bibinfo {author} {\bibfnamefont {H.}~\bibnamefont {Liao}}, \bibinfo
  {author} {\bibfnamefont {Z.}~\bibnamefont {Zhang}}, \bibinfo {author}
  {\bibfnamefont {S.}~\bibnamefont {Li}}, \bibinfo {author} {\bibfnamefont
  {L.}~\bibnamefont {Ling}}, \bibinfo {author} {\bibfnamefont {L.}~\bibnamefont
  {Zhang}}, \bibinfo {author} {\bibfnamefont {Y.}~\bibnamefont {Zou}}, \bibinfo
  {author} {\bibfnamefont {Z.}~\bibnamefont {Yang}}, \bibinfo {author}
  {\bibfnamefont {J.}~\bibnamefont {Wang}},\ and\ \bibinfo {author}
  {\bibfnamefont {Z.}~\bibnamefont {Wu}},\ }\href
  {https://doi.org/10.1038/srep16419} {\bibfield  {journal} {\bibinfo
  {journal} {Scientific Reports}\ }\textbf {\bibinfo {volume} {5}},\ \bibinfo
  {pages} {16419} (\bibinfo {year} {2015}{\natexlab{a}})}\BibitemShut {NoStop}%
\bibitem [{\citenamefont {Li}\ \emph {et~al.}(2015{\natexlab{b}})\citenamefont
  {Li}, \citenamefont {Chen}, \citenamefont {Tong}, \citenamefont {Pi},
  \citenamefont {Liu}, \citenamefont {Yang}, \citenamefont {Wang},\ and\
  \citenamefont {Zhang}}]{LiPRL2015}%
  \BibitemOpen
  \bibfield  {author} {\bibinfo {author} {\bibfnamefont {Y.}~\bibnamefont
  {Li}}, \bibinfo {author} {\bibfnamefont {G.}~\bibnamefont {Chen}}, \bibinfo
  {author} {\bibfnamefont {W.}~\bibnamefont {Tong}}, \bibinfo {author}
  {\bibfnamefont {L.}~\bibnamefont {Pi}}, \bibinfo {author} {\bibfnamefont
  {J.}~\bibnamefont {Liu}}, \bibinfo {author} {\bibfnamefont {Z.}~\bibnamefont
  {Yang}}, \bibinfo {author} {\bibfnamefont {X.}~\bibnamefont {Wang}},\ and\
  \bibinfo {author} {\bibfnamefont {Q.}~\bibnamefont {Zhang}},\ }\href
  {https://doi.org/10.1103/PhysRevLett.115.167203} {\bibfield  {journal}
  {\bibinfo  {journal} {Phys. Rev. Lett.}\ }\textbf {\bibinfo {volume} {115}},\
  \bibinfo {pages} {167203} (\bibinfo {year} {2015}{\natexlab{b}})}\BibitemShut
  {NoStop}%
\bibitem [{\citenamefont {Li}\ \emph {et~al.}(2016)\citenamefont {Li},
  \citenamefont {Adroja}, \citenamefont {Biswas}, \citenamefont {Baker},
  \citenamefont {Zhang}, \citenamefont {Liu}, \citenamefont {Tsirlin},
  \citenamefont {Gegenwart},\ and\ \citenamefont {Zhang}}]{Li_PRL2016}%
  \BibitemOpen
  \bibfield  {author} {\bibinfo {author} {\bibfnamefont {Y.}~\bibnamefont
  {Li}}, \bibinfo {author} {\bibfnamefont {D.}~\bibnamefont {Adroja}}, \bibinfo
  {author} {\bibfnamefont {P.~K.}\ \bibnamefont {Biswas}}, \bibinfo {author}
  {\bibfnamefont {P.~J.}\ \bibnamefont {Baker}}, \bibinfo {author}
  {\bibfnamefont {Q.}~\bibnamefont {Zhang}}, \bibinfo {author} {\bibfnamefont
  {J.}~\bibnamefont {Liu}}, \bibinfo {author} {\bibfnamefont {A.~A.}\
  \bibnamefont {Tsirlin}}, \bibinfo {author} {\bibfnamefont {P.}~\bibnamefont
  {Gegenwart}},\ and\ \bibinfo {author} {\bibfnamefont {Q.}~\bibnamefont
  {Zhang}},\ }\href {https://doi.org/10.1103/PhysRevLett.117.097201} {\bibfield
   {journal} {\bibinfo  {journal} {Phys. Rev. Lett.}\ }\textbf {\bibinfo
  {volume} {117}},\ \bibinfo {pages} {097201} (\bibinfo {year}
  {2016})}\BibitemShut {NoStop}%
\bibitem [{\citenamefont {Zhang}\ \emph {et~al.}(2018)\citenamefont {Zhang},
  \citenamefont {Mahmood}, \citenamefont {Daum}, \citenamefont {Dun},
  \citenamefont {Paddison}, \citenamefont {Laurita}, \citenamefont {Hong},
  \citenamefont {Zhou}, \citenamefont {Armitage},\ and\ \citenamefont
  {Mourigal}}]{ZhangPRX2018}%
  \BibitemOpen
  \bibfield  {author} {\bibinfo {author} {\bibfnamefont {X.}~\bibnamefont
  {Zhang}}, \bibinfo {author} {\bibfnamefont {F.}~\bibnamefont {Mahmood}},
  \bibinfo {author} {\bibfnamefont {M.}~\bibnamefont {Daum}}, \bibinfo {author}
  {\bibfnamefont {Z.}~\bibnamefont {Dun}}, \bibinfo {author} {\bibfnamefont
  {J.~A.~M.}\ \bibnamefont {Paddison}}, \bibinfo {author} {\bibfnamefont
  {N.~J.}\ \bibnamefont {Laurita}}, \bibinfo {author} {\bibfnamefont
  {T.}~\bibnamefont {Hong}}, \bibinfo {author} {\bibfnamefont {H.}~\bibnamefont
  {Zhou}}, \bibinfo {author} {\bibfnamefont {N.~P.}\ \bibnamefont {Armitage}},\
  and\ \bibinfo {author} {\bibfnamefont {M.}~\bibnamefont {Mourigal}},\ }\href
  {https://doi.org/10.1103/PhysRevX.8.031001} {\bibfield  {journal} {\bibinfo
  {journal} {Phys. Rev. X}\ }\textbf {\bibinfo {volume} {8}},\ \bibinfo {pages}
  {031001} (\bibinfo {year} {2018})}\BibitemShut {NoStop}%
\bibitem [{\citenamefont {Li}(2019)}]{Li_YMGOreview}%
  \BibitemOpen
  \bibfield  {author} {\bibinfo {author} {\bibfnamefont {Y.}~\bibnamefont
  {Li}},\ }\href {https://doi.org/https://doi.org/10.1002/qute.201900089}
  {\bibfield  {journal} {\bibinfo  {journal} {Advanced Quantum Technologies}\
  }\textbf {\bibinfo {volume} {2}},\ \bibinfo {pages} {1900089} (\bibinfo
  {year} {2019})},\ \Eprint
  {https://arxiv.org/abs/https://onlinelibrary.wiley.com/doi/pdf/10.1002/qute.201900089}
  {https://onlinelibrary.wiley.com/doi/pdf/10.1002/qute.201900089} \BibitemShut
  {NoStop}%
\bibitem [{\citenamefont {Ma}\ \emph {et~al.}(2018)\citenamefont {Ma},
  \citenamefont {Wang}, \citenamefont {Dong}, \citenamefont {Zhang},
  \citenamefont {Li}, \citenamefont {Zheng}, \citenamefont {Yu}, \citenamefont
  {Wang}, \citenamefont {Che}, \citenamefont {Ran}, \citenamefont {Bao},
  \citenamefont {Cai}, \citenamefont {\ifmmode~\check{C}\else
  \v{C}\fi{}erm\'ak}, \citenamefont {Schneidewind}, \citenamefont {Yano},
  \citenamefont {Gardner}, \citenamefont {Lu}, \citenamefont {Yu},
  \citenamefont {Liu}, \citenamefont {Li}, \citenamefont {Li},\ and\
  \citenamefont {Wen}}]{MaPRL2018}%
  \BibitemOpen
  \bibfield  {author} {\bibinfo {author} {\bibfnamefont {Z.}~\bibnamefont
  {Ma}}, \bibinfo {author} {\bibfnamefont {J.}~\bibnamefont {Wang}}, \bibinfo
  {author} {\bibfnamefont {Z.-Y.}\ \bibnamefont {Dong}}, \bibinfo {author}
  {\bibfnamefont {J.}~\bibnamefont {Zhang}}, \bibinfo {author} {\bibfnamefont
  {S.}~\bibnamefont {Li}}, \bibinfo {author} {\bibfnamefont {S.-H.}\
  \bibnamefont {Zheng}}, \bibinfo {author} {\bibfnamefont {Y.}~\bibnamefont
  {Yu}}, \bibinfo {author} {\bibfnamefont {W.}~\bibnamefont {Wang}}, \bibinfo
  {author} {\bibfnamefont {L.}~\bibnamefont {Che}}, \bibinfo {author}
  {\bibfnamefont {K.}~\bibnamefont {Ran}}, \bibinfo {author} {\bibfnamefont
  {S.}~\bibnamefont {Bao}}, \bibinfo {author} {\bibfnamefont {Z.}~\bibnamefont
  {Cai}}, \bibinfo {author} {\bibfnamefont {P.}~\bibnamefont
  {\ifmmode~\check{C}\else \v{C}\fi{}erm\'ak}}, \bibinfo {author}
  {\bibfnamefont {A.}~\bibnamefont {Schneidewind}}, \bibinfo {author}
  {\bibfnamefont {S.}~\bibnamefont {Yano}}, \bibinfo {author} {\bibfnamefont
  {J.~S.}\ \bibnamefont {Gardner}}, \bibinfo {author} {\bibfnamefont
  {X.}~\bibnamefont {Lu}}, \bibinfo {author} {\bibfnamefont {S.-L.}\
  \bibnamefont {Yu}}, \bibinfo {author} {\bibfnamefont {J.-M.}\ \bibnamefont
  {Liu}}, \bibinfo {author} {\bibfnamefont {S.}~\bibnamefont {Li}}, \bibinfo
  {author} {\bibfnamefont {J.-X.}\ \bibnamefont {Li}},\ and\ \bibinfo {author}
  {\bibfnamefont {J.}~\bibnamefont {Wen}},\ }\href
  {https://doi.org/10.1103/PhysRevLett.120.087201} {\bibfield  {journal}
  {\bibinfo  {journal} {Phys. Rev. Lett.}\ }\textbf {\bibinfo {volume} {120}},\
  \bibinfo {pages} {087201} (\bibinfo {year} {2018})}\BibitemShut {NoStop}%
\bibitem [{\citenamefont {Li}\ \emph {et~al.}(2017)\citenamefont {Li},
  \citenamefont {Adroja}, \citenamefont {Bewley}, \citenamefont {Voneshen},
  \citenamefont {Tsirlin}, \citenamefont {Gegenwart},\ and\ \citenamefont
  {Zhang}}]{LiPRL2017}%
  \BibitemOpen
  \bibfield  {author} {\bibinfo {author} {\bibfnamefont {Y.}~\bibnamefont
  {Li}}, \bibinfo {author} {\bibfnamefont {D.}~\bibnamefont {Adroja}}, \bibinfo
  {author} {\bibfnamefont {R.~I.}\ \bibnamefont {Bewley}}, \bibinfo {author}
  {\bibfnamefont {D.}~\bibnamefont {Voneshen}}, \bibinfo {author}
  {\bibfnamefont {A.~A.}\ \bibnamefont {Tsirlin}}, \bibinfo {author}
  {\bibfnamefont {P.}~\bibnamefont {Gegenwart}},\ and\ \bibinfo {author}
  {\bibfnamefont {Q.}~\bibnamefont {Zhang}},\ }\href
  {https://doi.org/10.1103/PhysRevLett.118.107202} {\bibfield  {journal}
  {\bibinfo  {journal} {Phys. Rev. Lett.}\ }\textbf {\bibinfo {volume} {118}},\
  \bibinfo {pages} {107202} (\bibinfo {year} {2017})}\BibitemShut {NoStop}%
\bibitem [{\citenamefont {Zhu}\ \emph {et~al.}(2017)\citenamefont {Zhu},
  \citenamefont {Maksimov}, \citenamefont {White},\ and\ \citenamefont
  {Chernyshev}}]{ZhuPRL2017}%
  \BibitemOpen
  \bibfield  {author} {\bibinfo {author} {\bibfnamefont {Z.}~\bibnamefont
  {Zhu}}, \bibinfo {author} {\bibfnamefont {P.~A.}\ \bibnamefont {Maksimov}},
  \bibinfo {author} {\bibfnamefont {S.~R.}\ \bibnamefont {White}},\ and\
  \bibinfo {author} {\bibfnamefont {A.~L.}\ \bibnamefont {Chernyshev}},\ }\href
  {https://doi.org/10.1103/PhysRevLett.119.157201} {\bibfield  {journal}
  {\bibinfo  {journal} {Phys. Rev. Lett.}\ }\textbf {\bibinfo {volume} {119}},\
  \bibinfo {pages} {157201} (\bibinfo {year} {2017})}\BibitemShut {NoStop}%
\bibitem [{\citenamefont {Parker}\ and\ \citenamefont
  {Balents}(2018)}]{ParkerPRB2018}%
  \BibitemOpen
  \bibfield  {author} {\bibinfo {author} {\bibfnamefont {E.}~\bibnamefont
  {Parker}}\ and\ \bibinfo {author} {\bibfnamefont {L.}~\bibnamefont
  {Balents}},\ }\href {https://doi.org/10.1103/PhysRevB.97.184413} {\bibfield
  {journal} {\bibinfo  {journal} {Phys. Rev. B}\ }\textbf {\bibinfo {volume}
  {97}},\ \bibinfo {pages} {184413} (\bibinfo {year} {2018})}\BibitemShut
  {NoStop}%
\bibitem [{\citenamefont {Ranjith}\ \emph {et~al.}(2019)\citenamefont
  {Ranjith}, \citenamefont {Luther}, \citenamefont {Reimann}, \citenamefont
  {Schmidt}, \citenamefont {Schlender}, \citenamefont {Sichelschmidt},
  \citenamefont {Yasuoka}, \citenamefont {Strydom}, \citenamefont {Skourski},
  \citenamefont {Wosnitza}, \citenamefont {K\"uhne}, \citenamefont {Doert},\
  and\ \citenamefont {Baenitz}}]{RanjithPRB2019}%
  \BibitemOpen
  \bibfield  {author} {\bibinfo {author} {\bibfnamefont {K.~M.}\ \bibnamefont
  {Ranjith}}, \bibinfo {author} {\bibfnamefont {S.}~\bibnamefont {Luther}},
  \bibinfo {author} {\bibfnamefont {T.}~\bibnamefont {Reimann}}, \bibinfo
  {author} {\bibfnamefont {B.}~\bibnamefont {Schmidt}}, \bibinfo {author}
  {\bibfnamefont {P.}~\bibnamefont {Schlender}}, \bibinfo {author}
  {\bibfnamefont {J.}~\bibnamefont {Sichelschmidt}}, \bibinfo {author}
  {\bibfnamefont {H.}~\bibnamefont {Yasuoka}}, \bibinfo {author} {\bibfnamefont
  {A.~M.}\ \bibnamefont {Strydom}}, \bibinfo {author} {\bibfnamefont
  {Y.}~\bibnamefont {Skourski}}, \bibinfo {author} {\bibfnamefont
  {J.}~\bibnamefont {Wosnitza}}, \bibinfo {author} {\bibfnamefont
  {H.}~\bibnamefont {K\"uhne}}, \bibinfo {author} {\bibfnamefont
  {T.}~\bibnamefont {Doert}},\ and\ \bibinfo {author} {\bibfnamefont
  {M.}~\bibnamefont {Baenitz}},\ }\href
  {https://doi.org/10.1103/PhysRevB.100.224417} {\bibfield  {journal} {\bibinfo
   {journal} {Phys. Rev. B}\ }\textbf {\bibinfo {volume} {100}},\ \bibinfo
  {pages} {224417} (\bibinfo {year} {2019})}\BibitemShut {NoStop}%
\bibitem [{\citenamefont {Bordelon}\ \emph {et~al.}(2019)\citenamefont
  {Bordelon}, \citenamefont {Kenney},\ and\ \citenamefont
  {Liu}}]{BordelonNature2019}%
  \BibitemOpen
  \bibfield  {author} {\bibinfo {author} {\bibfnamefont {M.}~\bibnamefont
  {Bordelon}}, \bibinfo {author} {\bibfnamefont {E.}~\bibnamefont {Kenney}},\
  and\ \bibinfo {author} {\bibfnamefont {C.}~\bibnamefont {Liu}},\ }\href
  {https://doi.org/10.1038/s41567-019-0594-5} {\bibfield  {journal} {\bibinfo
  {journal} {Nature Phys}\ }\textbf {\bibinfo {volume} {15}},\ \bibinfo {pages}
  {1058} (\bibinfo {year} {2019})}\BibitemShut {NoStop}%
\bibitem [{\citenamefont {Ding}\ \emph {et~al.}(2019)\citenamefont {Ding},
  \citenamefont {Manuel}, \citenamefont {Bachus}, \citenamefont {Gru\ss{}ler},
  \citenamefont {Gegenwart}, \citenamefont {Singleton}, \citenamefont
  {Johnson}, \citenamefont {Walker}, \citenamefont {Adroja}, \citenamefont
  {Hillier},\ and\ \citenamefont {Tsirlin}}]{DingPRB2019}%
  \BibitemOpen
  \bibfield  {author} {\bibinfo {author} {\bibfnamefont {L.}~\bibnamefont
  {Ding}}, \bibinfo {author} {\bibfnamefont {P.}~\bibnamefont {Manuel}},
  \bibinfo {author} {\bibfnamefont {S.}~\bibnamefont {Bachus}}, \bibinfo
  {author} {\bibfnamefont {F.}~\bibnamefont {Gru\ss{}ler}}, \bibinfo {author}
  {\bibfnamefont {P.}~\bibnamefont {Gegenwart}}, \bibinfo {author}
  {\bibfnamefont {J.}~\bibnamefont {Singleton}}, \bibinfo {author}
  {\bibfnamefont {R.~D.}\ \bibnamefont {Johnson}}, \bibinfo {author}
  {\bibfnamefont {H.~C.}\ \bibnamefont {Walker}}, \bibinfo {author}
  {\bibfnamefont {D.~T.}\ \bibnamefont {Adroja}}, \bibinfo {author}
  {\bibfnamefont {A.~D.}\ \bibnamefont {Hillier}},\ and\ \bibinfo {author}
  {\bibfnamefont {A.~A.}\ \bibnamefont {Tsirlin}},\ }\href
  {https://doi.org/10.1103/PhysRevB.100.144432} {\bibfield  {journal} {\bibinfo
   {journal} {Phys. Rev. B}\ }\textbf {\bibinfo {volume} {100}},\ \bibinfo
  {pages} {144432} (\bibinfo {year} {2019})}\BibitemShut {NoStop}%
\bibitem [{\citenamefont {Baenitz}\ \emph {et~al.}(2018)\citenamefont
  {Baenitz}, \citenamefont {Schlender}, \citenamefont {Sichelschmidt},
  \citenamefont {Onykiienko}, \citenamefont {Zangeneh}, \citenamefont
  {Ranjith}, \citenamefont {Sarkar}, \citenamefont {Hozoi}, \citenamefont
  {Walker}, \citenamefont {Orain}, \citenamefont {Yasuoka}, \citenamefont
  {van~den Brink}, \citenamefont {Klauss}, \citenamefont {Inosov},\ and\
  \citenamefont {Doert}}]{BaenitzPRB2018}%
  \BibitemOpen
  \bibfield  {author} {\bibinfo {author} {\bibfnamefont {M.}~\bibnamefont
  {Baenitz}}, \bibinfo {author} {\bibfnamefont {P.}~\bibnamefont {Schlender}},
  \bibinfo {author} {\bibfnamefont {J.}~\bibnamefont {Sichelschmidt}}, \bibinfo
  {author} {\bibfnamefont {Y.~A.}\ \bibnamefont {Onykiienko}}, \bibinfo
  {author} {\bibfnamefont {Z.}~\bibnamefont {Zangeneh}}, \bibinfo {author}
  {\bibfnamefont {K.~M.}\ \bibnamefont {Ranjith}}, \bibinfo {author}
  {\bibfnamefont {R.}~\bibnamefont {Sarkar}}, \bibinfo {author} {\bibfnamefont
  {L.}~\bibnamefont {Hozoi}}, \bibinfo {author} {\bibfnamefont {H.~C.}\
  \bibnamefont {Walker}}, \bibinfo {author} {\bibfnamefont {J.-C.}\
  \bibnamefont {Orain}}, \bibinfo {author} {\bibfnamefont {H.}~\bibnamefont
  {Yasuoka}}, \bibinfo {author} {\bibfnamefont {J.}~\bibnamefont {van~den
  Brink}}, \bibinfo {author} {\bibfnamefont {H.~H.}\ \bibnamefont {Klauss}},
  \bibinfo {author} {\bibfnamefont {D.~S.}\ \bibnamefont {Inosov}},\ and\
  \bibinfo {author} {\bibfnamefont {T.}~\bibnamefont {Doert}},\ }\href
  {https://doi.org/10.1103/PhysRevB.98.220409} {\bibfield  {journal} {\bibinfo
  {journal} {Phys. Rev. B}\ }\textbf {\bibinfo {volume} {98}},\ \bibinfo
  {pages} {220409} (\bibinfo {year} {2018})}\BibitemShut {NoStop}%
\bibitem [{\citenamefont {Guo}\ \emph {et~al.}(2020)\citenamefont {Guo},
  \citenamefont {Zhao}, \citenamefont {Ohira-Kawamura}, \citenamefont {Ling},
  \citenamefont {Wang}, \citenamefont {He}, \citenamefont {Nakajima},
  \citenamefont {Li},\ and\ \citenamefont {Zhang}}]{GuoPRM2020}%
  \BibitemOpen
  \bibfield  {author} {\bibinfo {author} {\bibfnamefont {J.}~\bibnamefont
  {Guo}}, \bibinfo {author} {\bibfnamefont {X.}~\bibnamefont {Zhao}}, \bibinfo
  {author} {\bibfnamefont {S.}~\bibnamefont {Ohira-Kawamura}}, \bibinfo
  {author} {\bibfnamefont {L.}~\bibnamefont {Ling}}, \bibinfo {author}
  {\bibfnamefont {J.}~\bibnamefont {Wang}}, \bibinfo {author} {\bibfnamefont
  {L.}~\bibnamefont {He}}, \bibinfo {author} {\bibfnamefont {K.}~\bibnamefont
  {Nakajima}}, \bibinfo {author} {\bibfnamefont {B.}~\bibnamefont {Li}},\ and\
  \bibinfo {author} {\bibfnamefont {Z.}~\bibnamefont {Zhang}},\ }\href
  {https://doi.org/10.1103/PhysRevMaterials.4.064410} {\bibfield  {journal}
  {\bibinfo  {journal} {Phys. Rev. Materials}\ }\textbf {\bibinfo {volume}
  {4}},\ \bibinfo {pages} {064410} (\bibinfo {year} {2020})}\BibitemShut
  {NoStop}%
\bibitem [{\citenamefont {Liu}\ \emph {et~al.}(2018)\citenamefont {Liu},
  \citenamefont {Zhang}, \citenamefont {Ji}, \citenamefont {Liu}, \citenamefont
  {Li}, \citenamefont {Wang}, \citenamefont {Lei}, \citenamefont {Chen},\ and\
  \citenamefont {Zhang}}]{LiuChinLett2018}%
  \BibitemOpen
  \bibfield  {author} {\bibinfo {author} {\bibfnamefont {W.}~\bibnamefont
  {Liu}}, \bibinfo {author} {\bibfnamefont {Z.}~\bibnamefont {Zhang}}, \bibinfo
  {author} {\bibfnamefont {J.}~\bibnamefont {Ji}}, \bibinfo {author}
  {\bibfnamefont {Y.}~\bibnamefont {Liu}}, \bibinfo {author} {\bibfnamefont
  {J.}~\bibnamefont {Li}}, \bibinfo {author} {\bibfnamefont {X.}~\bibnamefont
  {Wang}}, \bibinfo {author} {\bibfnamefont {H.}~\bibnamefont {Lei}}, \bibinfo
  {author} {\bibfnamefont {G.}~\bibnamefont {Chen}},\ and\ \bibinfo {author}
  {\bibfnamefont {Q.}~\bibnamefont {Zhang}},\ }\href
  {https://doi.org/10.1088/0256-307x/35/11/117501} {\bibfield  {journal}
  {\bibinfo  {journal} {Chinese Physics Letters}\ }\textbf {\bibinfo {volume}
  {35}},\ \bibinfo {pages} {117501} (\bibinfo {year} {2018})}\BibitemShut
  {NoStop}%
\bibitem [{\citenamefont {Bordelon}\ \emph {et~al.}(2020)\citenamefont
  {Bordelon}, \citenamefont {Liu}, \citenamefont {Posthuma}, \citenamefont
  {Sarte}, \citenamefont {Butch}, \citenamefont {Pajerowski}, \citenamefont
  {Banerjee}, \citenamefont {Balents},\ and\ \citenamefont
  {Wilson}}]{BordelonPRB2020}%
  \BibitemOpen
  \bibfield  {author} {\bibinfo {author} {\bibfnamefont {M.~M.}\ \bibnamefont
  {Bordelon}}, \bibinfo {author} {\bibfnamefont {C.}~\bibnamefont {Liu}},
  \bibinfo {author} {\bibfnamefont {L.}~\bibnamefont {Posthuma}}, \bibinfo
  {author} {\bibfnamefont {P.~M.}\ \bibnamefont {Sarte}}, \bibinfo {author}
  {\bibfnamefont {N.~P.}\ \bibnamefont {Butch}}, \bibinfo {author}
  {\bibfnamefont {D.~M.}\ \bibnamefont {Pajerowski}}, \bibinfo {author}
  {\bibfnamefont {A.}~\bibnamefont {Banerjee}}, \bibinfo {author}
  {\bibfnamefont {L.}~\bibnamefont {Balents}},\ and\ \bibinfo {author}
  {\bibfnamefont {S.~D.}\ \bibnamefont {Wilson}},\ }\href
  {https://doi.org/10.1103/PhysRevB.101.224427} {\bibfield  {journal} {\bibinfo
   {journal} {Phys. Rev. B}\ }\textbf {\bibinfo {volume} {101}},\ \bibinfo
  {pages} {224427} (\bibinfo {year} {2020})}\BibitemShut {NoStop}%
\bibitem [{\citenamefont {Guo}\ \emph {et~al.}(2019{\natexlab{a}})\citenamefont
  {Guo}, \citenamefont {Ghasemi}, \citenamefont {Broholm},\ and\ \citenamefont
  {Cava}}]{Guo_2019}%
  \BibitemOpen
  \bibfield  {author} {\bibinfo {author} {\bibfnamefont {S.}~\bibnamefont
  {Guo}}, \bibinfo {author} {\bibfnamefont {A.}~\bibnamefont {Ghasemi}},
  \bibinfo {author} {\bibfnamefont {C.~L.}\ \bibnamefont {Broholm}},\ and\
  \bibinfo {author} {\bibfnamefont {R.~J.}\ \bibnamefont {Cava}},\ }\href
  {https://doi.org/10.1103/PhysRevMaterials.3.094404} {\bibfield  {journal}
  {\bibinfo  {journal} {Phys. Rev. Materials}\ }\textbf {\bibinfo {volume}
  {3}},\ \bibinfo {pages} {094404} (\bibinfo {year}
  {2019}{\natexlab{a}})}\BibitemShut {NoStop}%
\bibitem [{\citenamefont {Sanders}\ \emph {et~al.}(2017)\citenamefont
  {Sanders}, \citenamefont {Cevallos},\ and\ \citenamefont
  {Cava}}]{Sanders_2017}%
  \BibitemOpen
  \bibfield  {author} {\bibinfo {author} {\bibfnamefont {M.~B.}\ \bibnamefont
  {Sanders}}, \bibinfo {author} {\bibfnamefont {F.~A.}\ \bibnamefont
  {Cevallos}},\ and\ \bibinfo {author} {\bibfnamefont {R.~J.}\ \bibnamefont
  {Cava}},\ }\href {https://doi.org/10.1088/2053-1591/aa60a2} {\bibfield
  {journal} {\bibinfo  {journal} {Materials Research Express}\ }\textbf
  {\bibinfo {volume} {4}},\ \bibinfo {pages} {036102} (\bibinfo {year}
  {2017})}\BibitemShut {NoStop}%
\bibitem [{\citenamefont {Svetlyakova}\ \emph {et~al.}(2013)\citenamefont
  {Svetlyakova}, \citenamefont {Kokh}, \citenamefont {Kononova}, \citenamefont
  {Fedorov}, \citenamefont {Rashchenko},\ and\ \citenamefont
  {Maillard}}]{Svetlyakova2013}%
  \BibitemOpen
  \bibfield  {author} {\bibinfo {author} {\bibfnamefont {T.~N.}\ \bibnamefont
  {Svetlyakova}}, \bibinfo {author} {\bibfnamefont {A.~E.}\ \bibnamefont
  {Kokh}}, \bibinfo {author} {\bibfnamefont {N.~G.}\ \bibnamefont {Kononova}},
  \bibinfo {author} {\bibfnamefont {P.~P.}\ \bibnamefont {Fedorov}}, \bibinfo
  {author} {\bibfnamefont {S.~V.}\ \bibnamefont {Rashchenko}},\ and\ \bibinfo
  {author} {\bibfnamefont {A.}~\bibnamefont {Maillard}},\ }\href
  {https://doi.org/10.1134/S1063774513010136} {\bibfield  {journal} {\bibinfo
  {journal} {Crystallography Reports}\ }\textbf {\bibinfo {volume} {58}},\
  \bibinfo {pages} {54} (\bibinfo {year} {2013})}\BibitemShut {NoStop}%
\bibitem [{\citenamefont {Pan}\ \emph {et~al.}(2021)\citenamefont {Pan},
  \citenamefont {Ni}, \citenamefont {He}, \citenamefont {Yu}, \citenamefont
  {Xu},\ and\ \citenamefont {Li}}]{Pan_PRB}%
  \BibitemOpen
  \bibfield  {author} {\bibinfo {author} {\bibfnamefont {B.~L.}\ \bibnamefont
  {Pan}}, \bibinfo {author} {\bibfnamefont {J.~M.}\ \bibnamefont {Ni}},
  \bibinfo {author} {\bibfnamefont {L.~P.}\ \bibnamefont {He}}, \bibinfo
  {author} {\bibfnamefont {Y.~J.}\ \bibnamefont {Yu}}, \bibinfo {author}
  {\bibfnamefont {Y.}~\bibnamefont {Xu}},\ and\ \bibinfo {author}
  {\bibfnamefont {S.~Y.}\ \bibnamefont {Li}},\ }\href
  {https://doi.org/10.1103/PhysRevB.103.104412} {\bibfield  {journal} {\bibinfo
   {journal} {Phys. Rev. B}\ }\textbf {\bibinfo {volume} {103}},\ \bibinfo
  {pages} {104412} (\bibinfo {year} {2021})}\BibitemShut {NoStop}%
\bibitem [{\citenamefont {Kuznetsov}\ \emph {et~al.}(2022)\citenamefont
  {Kuznetsov}, \citenamefont {Kokh}, \citenamefont {Sagatov}, \citenamefont
  {Gavryushkin}, \citenamefont {Molokeev}, \citenamefont {Svetlichnyi},
  \citenamefont {Lapin}, \citenamefont {Kononova}, \citenamefont {Shevchenko},
  \citenamefont {Bolatov}, \citenamefont {Uralbekov}, \citenamefont
  {Goreiavcheva},\ and\ \citenamefont {Kokh}}]{Kuznetsov_NSYBO}%
  \BibitemOpen
  \bibfield  {author} {\bibinfo {author} {\bibfnamefont {A.~B.}\ \bibnamefont
  {Kuznetsov}}, \bibinfo {author} {\bibfnamefont {K.~A.}\ \bibnamefont {Kokh}},
  \bibinfo {author} {\bibfnamefont {N.}~\bibnamefont {Sagatov}}, \bibinfo
  {author} {\bibfnamefont {P.~N.}\ \bibnamefont {Gavryushkin}}, \bibinfo
  {author} {\bibfnamefont {M.~S.}\ \bibnamefont {Molokeev}}, \bibinfo {author}
  {\bibfnamefont {V.~A.}\ \bibnamefont {Svetlichnyi}}, \bibinfo {author}
  {\bibfnamefont {I.~N.}\ \bibnamefont {Lapin}}, \bibinfo {author}
  {\bibfnamefont {N.~G.}\ \bibnamefont {Kononova}}, \bibinfo {author}
  {\bibfnamefont {V.~S.}\ \bibnamefont {Shevchenko}}, \bibinfo {author}
  {\bibfnamefont {A.}~\bibnamefont {Bolatov}}, \bibinfo {author} {\bibfnamefont
  {B.}~\bibnamefont {Uralbekov}}, \bibinfo {author} {\bibfnamefont {A.~A.}\
  \bibnamefont {Goreiavcheva}},\ and\ \bibinfo {author} {\bibfnamefont {A.~E.}\
  \bibnamefont {Kokh}},\ }\href {https://doi.org/10.1021/acs.inorgchem.2c00596}
  {\bibfield  {journal} {\bibinfo  {journal} {Inorganic Chemistry}\ }\textbf
  {\bibinfo {volume} {61}},\ \bibinfo {pages} {7497} (\bibinfo {year}
  {2022})},\ \bibinfo {note} {pMID: 35503917},\ \Eprint
  {https://arxiv.org/abs/https://doi.org/10.1021/acs.inorgchem.2c00596}
  {https://doi.org/10.1021/acs.inorgchem.2c00596} \BibitemShut {NoStop}%
\bibitem [{\citenamefont {Bag}\ \emph {et~al.}(2021)\citenamefont {Bag},
  \citenamefont {Ennis}, \citenamefont {Liu}, \citenamefont {Dissanayake},
  \citenamefont {Shi}, \citenamefont {Liu}, \citenamefont {Balents},\ and\
  \citenamefont {Haravifard}}]{BYBOPRB}%
  \BibitemOpen
  \bibfield  {author} {\bibinfo {author} {\bibfnamefont {R.}~\bibnamefont
  {Bag}}, \bibinfo {author} {\bibfnamefont {M.}~\bibnamefont {Ennis}}, \bibinfo
  {author} {\bibfnamefont {C.}~\bibnamefont {Liu}}, \bibinfo {author}
  {\bibfnamefont {S.~E.}\ \bibnamefont {Dissanayake}}, \bibinfo {author}
  {\bibfnamefont {Z.}~\bibnamefont {Shi}}, \bibinfo {author} {\bibfnamefont
  {J.}~\bibnamefont {Liu}}, \bibinfo {author} {\bibfnamefont {L.}~\bibnamefont
  {Balents}},\ and\ \bibinfo {author} {\bibfnamefont {S.}~\bibnamefont
  {Haravifard}},\ }\href {https://doi.org/10.1103/PhysRevB.104.L220403}
  {\bibfield  {journal} {\bibinfo  {journal} {Phys. Rev. B}\ }\textbf {\bibinfo
  {volume} {104}},\ \bibinfo {pages} {L220403} (\bibinfo {year}
  {2021})}\BibitemShut {NoStop}%
\bibitem [{\citenamefont {Cevallos}\ \emph
  {et~al.}(2018{\natexlab{a}})\citenamefont {Cevallos}, \citenamefont
  {Stolze},\ and\ \citenamefont {Cava}}]{Cevallos_SSC2018}%
  \BibitemOpen
  \bibfield  {author} {\bibinfo {author} {\bibfnamefont {F.~A.}\ \bibnamefont
  {Cevallos}}, \bibinfo {author} {\bibfnamefont {K.}~\bibnamefont {Stolze}},\
  and\ \bibinfo {author} {\bibfnamefont {R.~J.}\ \bibnamefont {Cava}},\ }\href
  {https://doi.org/https://doi.org/10.1016/j.ssc.2018.03.015} {\bibfield
  {journal} {\bibinfo  {journal} {Solid State Communications}\ }\textbf
  {\bibinfo {volume} {276}},\ \bibinfo {pages} {5} (\bibinfo {year}
  {2018}{\natexlab{a}})}\BibitemShut {NoStop}%
\bibitem [{\citenamefont {Cevallos}\ \emph
  {et~al.}(2018{\natexlab{b}})\citenamefont {Cevallos}, \citenamefont {Stolze},
  \citenamefont {Kong},\ and\ \citenamefont {Cava}}]{Cevallos_TmMgGaO4}%
  \BibitemOpen
  \bibfield  {author} {\bibinfo {author} {\bibfnamefont {F.~A.}\ \bibnamefont
  {Cevallos}}, \bibinfo {author} {\bibfnamefont {K.}~\bibnamefont {Stolze}},
  \bibinfo {author} {\bibfnamefont {T.}~\bibnamefont {Kong}},\ and\ \bibinfo
  {author} {\bibfnamefont {R.}~\bibnamefont {Cava}},\ }\href
  {https://doi.org/https://doi.org/10.1016/j.materresbull.2018.04.042}
  {\bibfield  {journal} {\bibinfo  {journal} {Materials Research Bulletin}\
  }\textbf {\bibinfo {volume} {105}},\ \bibinfo {pages} {154} (\bibinfo {year}
  {2018}{\natexlab{b}})}\BibitemShut {NoStop}%
\bibitem [{\citenamefont {Hashimoto}\ \emph {et~al.}(2003)\citenamefont
  {Hashimoto}, \citenamefont {Wakeshima},\ and\ \citenamefont
  {Hinatsu}}]{Hashimoto_2003}%
  \BibitemOpen
  \bibfield  {author} {\bibinfo {author} {\bibfnamefont {Y.}~\bibnamefont
  {Hashimoto}}, \bibinfo {author} {\bibfnamefont {M.}~\bibnamefont
  {Wakeshima}},\ and\ \bibinfo {author} {\bibfnamefont {Y.}~\bibnamefont
  {Hinatsu}},\ }\href
  {https://doi.org/https://doi.org/10.1016/j.jssc.2003.08.001} {\bibfield
  {journal} {\bibinfo  {journal} {Journal of Solid State Chemistry}\ }\textbf
  {\bibinfo {volume} {176}},\ \bibinfo {pages} {266} (\bibinfo {year}
  {2003})}\BibitemShut {NoStop}%
\bibitem [{\citenamefont {Xing}\ \emph {et~al.}(2019)\citenamefont {Xing},
  \citenamefont {Sanjeewa}, \citenamefont {Kim}, \citenamefont {Meier},
  \citenamefont {May}, \citenamefont {Zheng}, \citenamefont {Custelcean},
  \citenamefont {Stewart},\ and\ \citenamefont {Sefat}}]{XingPRM2019}%
  \BibitemOpen
  \bibfield  {author} {\bibinfo {author} {\bibfnamefont {J.}~\bibnamefont
  {Xing}}, \bibinfo {author} {\bibfnamefont {L.~D.}\ \bibnamefont {Sanjeewa}},
  \bibinfo {author} {\bibfnamefont {J.}~\bibnamefont {Kim}}, \bibinfo {author}
  {\bibfnamefont {W.~R.}\ \bibnamefont {Meier}}, \bibinfo {author}
  {\bibfnamefont {A.~F.}\ \bibnamefont {May}}, \bibinfo {author} {\bibfnamefont
  {Q.}~\bibnamefont {Zheng}}, \bibinfo {author} {\bibfnamefont
  {R.}~\bibnamefont {Custelcean}}, \bibinfo {author} {\bibfnamefont {G.~R.}\
  \bibnamefont {Stewart}},\ and\ \bibinfo {author} {\bibfnamefont {A.~S.}\
  \bibnamefont {Sefat}},\ }\href
  {https://doi.org/10.1103/PhysRevMaterials.3.114413} {\bibfield  {journal}
  {\bibinfo  {journal} {Phys. Rev. Materials}\ }\textbf {\bibinfo {volume}
  {3}},\ \bibinfo {pages} {114413} (\bibinfo {year} {2019})}\BibitemShut
  {NoStop}%
\bibitem [{\citenamefont {Deng}\ \emph {et~al.}(2002)\citenamefont {Deng},
  \citenamefont {Ellis},\ and\ \citenamefont {Ibers}}]{Deng_RbLnSe2}%
  \BibitemOpen
  \bibfield  {author} {\bibinfo {author} {\bibfnamefont {B.}~\bibnamefont
  {Deng}}, \bibinfo {author} {\bibfnamefont {D.~E.}\ \bibnamefont {Ellis}},\
  and\ \bibinfo {author} {\bibfnamefont {J.~A.}\ \bibnamefont {Ibers}},\ }\href
  {https://doi.org/10.1021/ic020324j} {\bibfield  {journal} {\bibinfo
  {journal} {Inorganic Chemistry}\ }\textbf {\bibinfo {volume} {41}},\ \bibinfo
  {pages} {5716} (\bibinfo {year} {2002})},\ \bibinfo {note} {pMID: 12401076},\
  \Eprint {https://arxiv.org/abs/https://doi.org/10.1021/ic020324j}
  {https://doi.org/10.1021/ic020324j} \BibitemShut {NoStop}%
\bibitem [{\citenamefont {Sanjeewa}\ \emph {et~al.}(2022)\citenamefont
  {Sanjeewa}, \citenamefont {Xing}, \citenamefont {Taddei},\ and\ \citenamefont
  {Sefat}}]{Sanjeewa_JSSC2022}%
  \BibitemOpen
  \bibfield  {author} {\bibinfo {author} {\bibfnamefont {L.~D.}\ \bibnamefont
  {Sanjeewa}}, \bibinfo {author} {\bibfnamefont {J.}~\bibnamefont {Xing}},
  \bibinfo {author} {\bibfnamefont {K.~M.}\ \bibnamefont {Taddei}},\ and\
  \bibinfo {author} {\bibfnamefont {A.~S.}\ \bibnamefont {Sefat}},\ }\href
  {https://doi.org/https://doi.org/10.1016/j.jssc.2022.122917} {\bibfield
  {journal} {\bibinfo  {journal} {Journal of Solid State Chemistry}\ }\textbf
  {\bibinfo {volume} {308}},\ \bibinfo {pages} {122917} (\bibinfo {year}
  {2022})}\BibitemShut {NoStop}%
\bibitem [{\citenamefont {Havl{\'{a}}k}\ \emph {et~al.}(2015)\citenamefont
  {Havl{\'{a}}k}, \citenamefont {F{\'{a}}bry}, \citenamefont {Henriques},\ and\
  \citenamefont {Du{\v{s}}ek}}]{Havlak_2015}%
  \BibitemOpen
  \bibfield  {author} {\bibinfo {author} {\bibfnamefont {L.}~\bibnamefont
  {Havl{\'{a}}k}}, \bibinfo {author} {\bibfnamefont {J.}~\bibnamefont
  {F{\'{a}}bry}}, \bibinfo {author} {\bibfnamefont {M.}~\bibnamefont
  {Henriques}},\ and\ \bibinfo {author} {\bibfnamefont {M.}~\bibnamefont
  {Du{\v{s}}ek}},\ }\href {https://doi.org/10.1107/S2053229615011833}
  {\bibfield  {journal} {\bibinfo  {journal} {Acta Crystallographica Section
  C}\ }\textbf {\bibinfo {volume} {71}},\ \bibinfo {pages} {623} (\bibinfo
  {year} {2015})}\BibitemShut {NoStop}%
\bibitem [{\citenamefont {Dong}\ \emph {et~al.}(2008)\citenamefont {Dong},
  \citenamefont {Doi},\ and\ \citenamefont {Hinatsu}}]{Dong_2008}%
  \BibitemOpen
  \bibfield  {author} {\bibinfo {author} {\bibfnamefont {B.}~\bibnamefont
  {Dong}}, \bibinfo {author} {\bibfnamefont {Y.}~\bibnamefont {Doi}},\ and\
  \bibinfo {author} {\bibfnamefont {Y.}~\bibnamefont {Hinatsu}},\ }\href
  {https://doi.org/https://doi.org/10.1016/j.jallcom.2006.11.053} {\bibfield
  {journal} {\bibinfo  {journal} {Journal of Alloys and Compounds}\ }\textbf
  {\bibinfo {volume} {453}},\ \bibinfo {pages} {282} (\bibinfo {year}
  {2008})}\BibitemShut {NoStop}%
\bibitem [{\citenamefont {Kononova}\ \emph {et~al.}(2016)\citenamefont
  {Kononova}, \citenamefont {Shevchenko}, \citenamefont {Kokh}, \citenamefont
  {Nabeeva}, \citenamefont {Chapron}, \citenamefont {Maillard}, \citenamefont
  {Bolatov},\ and\ \citenamefont {Uralbekov}}]{Kononova_2016}%
  \BibitemOpen
  \bibfield  {author} {\bibinfo {author} {\bibfnamefont {N.}~\bibnamefont
  {Kononova}}, \bibinfo {author} {\bibfnamefont {V.}~\bibnamefont
  {Shevchenko}}, \bibinfo {author} {\bibfnamefont {A.}~\bibnamefont {Kokh}},
  \bibinfo {author} {\bibfnamefont {T.}~\bibnamefont {Nabeeva}}, \bibinfo
  {author} {\bibfnamefont {D.}~\bibnamefont {Chapron}}, \bibinfo {author}
  {\bibfnamefont {A.}~\bibnamefont {Maillard}}, \bibinfo {author}
  {\bibfnamefont {A.}~\bibnamefont {Bolatov}},\ and\ \bibinfo {author}
  {\bibfnamefont {B.}~\bibnamefont {Uralbekov}},\ }\href
  {https://doi.org/10.1590/1980-5373-mr-2016-0081} {\bibfield  {journal}
  {\bibinfo  {journal} {Materials Research}\ }\textbf {\bibinfo {volume}
  {19}},\ \bibinfo {pages} {834} (\bibinfo {year} {2016})}\BibitemShut
  {NoStop}%
\bibitem [{\citenamefont {Guo}\ \emph {et~al.}(2019{\natexlab{b}})\citenamefont
  {Guo}, \citenamefont {Kong},\ and\ \citenamefont {Cava}}]{Guo_MRE2019}%
  \BibitemOpen
  \bibfield  {author} {\bibinfo {author} {\bibfnamefont {S.}~\bibnamefont
  {Guo}}, \bibinfo {author} {\bibfnamefont {T.}~\bibnamefont {Kong}},\ and\
  \bibinfo {author} {\bibfnamefont {R.~J.}\ \bibnamefont {Cava}},\ }\href
  {https://doi.org/10.1088/2053-1591/ab3d8e} {\bibfield  {journal} {\bibinfo
  {journal} {Materials Research Express}\ }\textbf {\bibinfo {volume} {6}},\
  \bibinfo {pages} {106110} (\bibinfo {year} {2019}{\natexlab{b}})}\BibitemShut
  {NoStop}%
\bibitem [{\citenamefont {Guo}\ \emph {et~al.}(2019{\natexlab{c}})\citenamefont
  {Guo}, \citenamefont {Kong}, \citenamefont {Xie}, \citenamefont {Nguyen},
  \citenamefont {Stolze}, \citenamefont {Cevallos},\ and\ \citenamefont
  {Cava}}]{Guo_InorgChem}%
  \BibitemOpen
  \bibfield  {author} {\bibinfo {author} {\bibfnamefont {S.}~\bibnamefont
  {Guo}}, \bibinfo {author} {\bibfnamefont {T.}~\bibnamefont {Kong}}, \bibinfo
  {author} {\bibfnamefont {W.}~\bibnamefont {Xie}}, \bibinfo {author}
  {\bibfnamefont {L.}~\bibnamefont {Nguyen}}, \bibinfo {author} {\bibfnamefont
  {K.}~\bibnamefont {Stolze}}, \bibinfo {author} {\bibfnamefont {F.~A.}\
  \bibnamefont {Cevallos}},\ and\ \bibinfo {author} {\bibfnamefont {R.~J.}\
  \bibnamefont {Cava}},\ }\href {https://doi.org/10.1021/acs.inorgchem.8b03372}
  {\bibfield  {journal} {\bibinfo  {journal} {Inorganic Chemistry}\ }\textbf
  {\bibinfo {volume} {58}},\ \bibinfo {pages} {3308} (\bibinfo {year}
  {2019}{\natexlab{c}})},\ \Eprint
  {https://arxiv.org/abs/https://doi.org/10.1021/acs.inorgchem.8b03372}
  {https://doi.org/10.1021/acs.inorgchem.8b03372} \BibitemShut {NoStop}%
\bibitem [{\citenamefont {Cai}\ \emph {et~al.}(2020)\citenamefont {Cai},
  \citenamefont {Lygouras}, \citenamefont {Thomas}, \citenamefont {Wilson},
  \citenamefont {Beare}, \citenamefont {Sharma}, \citenamefont {Marjerrison},
  \citenamefont {Yahne}, \citenamefont {Ross}, \citenamefont {Gong},
  \citenamefont {Uemura}, \citenamefont {Dabkowska},\ and\ \citenamefont
  {Luke}}]{Cai_muSR2020}%
  \BibitemOpen
  \bibfield  {author} {\bibinfo {author} {\bibfnamefont {Y.}~\bibnamefont
  {Cai}}, \bibinfo {author} {\bibfnamefont {C.}~\bibnamefont {Lygouras}},
  \bibinfo {author} {\bibfnamefont {G.}~\bibnamefont {Thomas}}, \bibinfo
  {author} {\bibfnamefont {M.~N.}\ \bibnamefont {Wilson}}, \bibinfo {author}
  {\bibfnamefont {J.}~\bibnamefont {Beare}}, \bibinfo {author} {\bibfnamefont
  {S.}~\bibnamefont {Sharma}}, \bibinfo {author} {\bibfnamefont {C.~A.}\
  \bibnamefont {Marjerrison}}, \bibinfo {author} {\bibfnamefont {D.~R.}\
  \bibnamefont {Yahne}}, \bibinfo {author} {\bibfnamefont {K.~A.}\ \bibnamefont
  {Ross}}, \bibinfo {author} {\bibfnamefont {Z.}~\bibnamefont {Gong}}, \bibinfo
  {author} {\bibfnamefont {Y.~J.}\ \bibnamefont {Uemura}}, \bibinfo {author}
  {\bibfnamefont {H.~A.}\ \bibnamefont {Dabkowska}},\ and\ \bibinfo {author}
  {\bibfnamefont {G.~M.}\ \bibnamefont {Luke}},\ }\href
  {https://doi.org/10.1103/PhysRevB.101.094432} {\bibfield  {journal} {\bibinfo
   {journal} {Phys. Rev. B}\ }\textbf {\bibinfo {volume} {101}},\ \bibinfo
  {pages} {094432} (\bibinfo {year} {2020})}\BibitemShut {NoStop}%
\bibitem [{\citenamefont {Gao}\ \emph {et~al.}(2020)\citenamefont {Gao},
  \citenamefont {Xiao}, \citenamefont {Kamazawa}, \citenamefont {Ikeuchi},
  \citenamefont {Biner}, \citenamefont {Kr\"amer}, \citenamefont {R\"uegg},\
  and\ \citenamefont {Arima}}]{Gao_2020}%
  \BibitemOpen
  \bibfield  {author} {\bibinfo {author} {\bibfnamefont {S.}~\bibnamefont
  {Gao}}, \bibinfo {author} {\bibfnamefont {F.}~\bibnamefont {Xiao}}, \bibinfo
  {author} {\bibfnamefont {K.}~\bibnamefont {Kamazawa}}, \bibinfo {author}
  {\bibfnamefont {K.}~\bibnamefont {Ikeuchi}}, \bibinfo {author} {\bibfnamefont
  {D.}~\bibnamefont {Biner}}, \bibinfo {author} {\bibfnamefont {K.~W.}\
  \bibnamefont {Kr\"amer}}, \bibinfo {author} {\bibfnamefont {C.}~\bibnamefont
  {R\"uegg}},\ and\ \bibinfo {author} {\bibfnamefont {T.-h.}\ \bibnamefont
  {Arima}},\ }\href {https://doi.org/10.1103/PhysRevB.102.024424} {\bibfield
  {journal} {\bibinfo  {journal} {Phys. Rev. B}\ }\textbf {\bibinfo {volume}
  {102}},\ \bibinfo {pages} {024424} (\bibinfo {year} {2020})}\BibitemShut
  {NoStop}%
\bibitem [{\citenamefont {Scheie}\ \emph {et~al.}(2020)\citenamefont {Scheie},
  \citenamefont {Garlea}, \citenamefont {Sanjeewa}, \citenamefont {Xing},\ and\
  \citenamefont {Sefat}}]{Scheie_2020}%
  \BibitemOpen
  \bibfield  {author} {\bibinfo {author} {\bibfnamefont {A.}~\bibnamefont
  {Scheie}}, \bibinfo {author} {\bibfnamefont {V.~O.}\ \bibnamefont {Garlea}},
  \bibinfo {author} {\bibfnamefont {L.~D.}\ \bibnamefont {Sanjeewa}}, \bibinfo
  {author} {\bibfnamefont {J.}~\bibnamefont {Xing}},\ and\ \bibinfo {author}
  {\bibfnamefont {A.~S.}\ \bibnamefont {Sefat}},\ }\href
  {https://doi.org/10.1103/PhysRevB.101.144432} {\bibfield  {journal} {\bibinfo
   {journal} {Phys. Rev. B}\ }\textbf {\bibinfo {volume} {101}},\ \bibinfo
  {pages} {144432} (\bibinfo {year} {2020})}\BibitemShut {NoStop}%
\bibitem [{\citenamefont {Xing}\ \emph {et~al.}(2021)\citenamefont {Xing},
  \citenamefont {Taddei}, \citenamefont {Sanjeewa}, \citenamefont {Fishman},
  \citenamefont {Daum}, \citenamefont {Mourigal}, \citenamefont {dela Cruz},\
  and\ \citenamefont {Sefat}}]{Xing_2021}%
  \BibitemOpen
  \bibfield  {author} {\bibinfo {author} {\bibfnamefont {J.}~\bibnamefont
  {Xing}}, \bibinfo {author} {\bibfnamefont {K.~M.}\ \bibnamefont {Taddei}},
  \bibinfo {author} {\bibfnamefont {L.~D.}\ \bibnamefont {Sanjeewa}}, \bibinfo
  {author} {\bibfnamefont {R.~S.}\ \bibnamefont {Fishman}}, \bibinfo {author}
  {\bibfnamefont {M.}~\bibnamefont {Daum}}, \bibinfo {author} {\bibfnamefont
  {M.}~\bibnamefont {Mourigal}}, \bibinfo {author} {\bibfnamefont
  {C.}~\bibnamefont {dela Cruz}},\ and\ \bibinfo {author} {\bibfnamefont
  {A.~S.}\ \bibnamefont {Sefat}},\ }\href
  {https://doi.org/10.1103/PhysRevB.103.144413} {\bibfield  {journal} {\bibinfo
   {journal} {Phys. Rev. B}\ }\textbf {\bibinfo {volume} {103}},\ \bibinfo
  {pages} {144413} (\bibinfo {year} {2021})}\BibitemShut {NoStop}%
\bibitem [{\citenamefont {Ding}\ \emph {et~al.}(2023)\citenamefont {Ding},
  \citenamefont {Wo}, \citenamefont {Luo}, \citenamefont {Gu}, \citenamefont
  {Gu}, \citenamefont {Bewley}, \citenamefont {Chen},\ and\ \citenamefont
  {Zhao}}]{Ding_2023}%
  \BibitemOpen
  \bibfield  {author} {\bibinfo {author} {\bibfnamefont {G.}~\bibnamefont
  {Ding}}, \bibinfo {author} {\bibfnamefont {H.}~\bibnamefont {Wo}}, \bibinfo
  {author} {\bibfnamefont {R.~L.}\ \bibnamefont {Luo}}, \bibinfo {author}
  {\bibfnamefont {Y.}~\bibnamefont {Gu}}, \bibinfo {author} {\bibfnamefont
  {Y.}~\bibnamefont {Gu}}, \bibinfo {author} {\bibfnamefont {R.}~\bibnamefont
  {Bewley}}, \bibinfo {author} {\bibfnamefont {G.}~\bibnamefont {Chen}},\ and\
  \bibinfo {author} {\bibfnamefont {J.}~\bibnamefont {Zhao}},\ }\href
  {https://doi.org/10.1103/PhysRevB.107.L100411} {\bibfield  {journal}
  {\bibinfo  {journal} {Phys. Rev. B}\ }\textbf {\bibinfo {volume} {107}},\
  \bibinfo {pages} {L100411} (\bibinfo {year} {2023})}\BibitemShut {NoStop}%
\bibitem [{\citenamefont {Granroth}\ \emph {et~al.}(2010)\citenamefont
  {Granroth}, \citenamefont {Kolesnikov}, \citenamefont {Sherline},
  \citenamefont {Clancy}, \citenamefont {Ross}, \citenamefont {Ruff},
  \citenamefont {Gaulin},\ and\ \citenamefont {Nagler}}]{SEQUOIA}%
  \BibitemOpen
  \bibfield  {author} {\bibinfo {author} {\bibfnamefont {G.~E.}\ \bibnamefont
  {Granroth}}, \bibinfo {author} {\bibfnamefont {A.~I.}\ \bibnamefont
  {Kolesnikov}}, \bibinfo {author} {\bibfnamefont {T.~E.}\ \bibnamefont
  {Sherline}}, \bibinfo {author} {\bibfnamefont {J.~P.}\ \bibnamefont
  {Clancy}}, \bibinfo {author} {\bibfnamefont {K.~A.}\ \bibnamefont {Ross}},
  \bibinfo {author} {\bibfnamefont {J.~P.~C.}\ \bibnamefont {Ruff}}, \bibinfo
  {author} {\bibfnamefont {B.~D.}\ \bibnamefont {Gaulin}},\ and\ \bibinfo
  {author} {\bibfnamefont {S.~E.}\ \bibnamefont {Nagler}},\ }\href
  {https://doi.org/10.1088/1742-6596/251/1/012058} {\bibfield  {journal}
  {\bibinfo  {journal} {Journal of Physics: Conference Series}\ }\textbf
  {\bibinfo {volume} {251}},\ \bibinfo {pages} {012058} (\bibinfo {year}
  {2010})}\BibitemShut {NoStop}%
\bibitem [{\citenamefont {Dimeo}\ \emph {et~al.}(2009)\citenamefont {Dimeo},
  \citenamefont {Azuah}, \citenamefont {Kneller}, \citenamefont {Qiu},
  \citenamefont {Tregenna-Piggott}, \citenamefont {Brown},\ and\ \citenamefont
  {Copley}}]{DAVE}%
  \BibitemOpen
  \bibfield  {author} {\bibinfo {author} {\bibfnamefont {R.}~\bibnamefont
  {Dimeo}}, \bibinfo {author} {\bibfnamefont {R.}~\bibnamefont {Azuah}},
  \bibinfo {author} {\bibfnamefont {L.}~\bibnamefont {Kneller}}, \bibinfo
  {author} {\bibfnamefont {Y.}~\bibnamefont {Qiu}}, \bibinfo {author}
  {\bibfnamefont {P.}~\bibnamefont {Tregenna-Piggott}}, \bibinfo {author}
  {\bibfnamefont {C.}~\bibnamefont {Brown}},\ and\ \bibinfo {author}
  {\bibfnamefont {J.}~\bibnamefont {Copley}},\ }\href
  {https://doi.org/10.6028/jres.114.025} {\bibfield  {journal} {\bibinfo
  {journal} {Journal of research of the National Institute of Standards and
  Technology}\ }\textbf {\bibinfo {volume} {114}},\ \bibinfo {pages} {341}
  (\bibinfo {year} {2009})}\BibitemShut {NoStop}%
\bibitem [{\citenamefont {Scheie}(2021)}]{PyCrystalField}%
  \BibitemOpen
  \bibfield  {author} {\bibinfo {author} {\bibfnamefont {A.}~\bibnamefont
  {Scheie}},\ }\href {https://doi.org/10.1107/S160057672001554X} {\bibfield
  {journal} {\bibinfo  {journal} {Journal of Applied Crystallography}\ }\textbf
  {\bibinfo {volume} {54}},\ \bibinfo {pages} {356} (\bibinfo {year}
  {2021})}\BibitemShut {NoStop}%
\bibitem [{beb()}]{bebo_lattice_data}%
  \BibitemOpen
  \href@noop {} {}\bibinfo {note} {{Lattice data at
  \url{https://materialsproject.org/materials/mp-1214885/}.}}\BibitemShut
  {Stop}%
\bibitem [{\citenamefont {Ilyukhin}(1993)}]{Ilyukhin1993}%
  \BibitemOpen
  \bibfield  {author} {\bibinfo {author} {\bibfnamefont {A.}~\bibnamefont
  {Ilyukhin}},\ }\href@noop {} {\bibfield  {journal} {\bibinfo  {journal}
  {Russ. J. Inorg. Chem.}\ }\textbf {\bibinfo {volume} {38}},\ \bibinfo {pages}
  {1625 } (\bibinfo {year} {1993})}\BibitemShut {NoStop}%
\bibitem [{\citenamefont {Khamaganova}\ \emph {et~al.}(1999)\citenamefont
  {Khamaganova}, \citenamefont {Kuperman},\ and\ \citenamefont
  {Bazarova}}]{Khamaganova1999}%
  \BibitemOpen
  \bibfield  {author} {\bibinfo {author} {\bibfnamefont {T.}~\bibnamefont
  {Khamaganova}}, \bibinfo {author} {\bibfnamefont {N.}~\bibnamefont
  {Kuperman}},\ and\ \bibinfo {author} {\bibfnamefont {Z.}~\bibnamefont
  {Bazarova}},\ }\href {https://doi.org/https://doi.org/10.1006/jssc.1999.8163}
  {\bibfield  {journal} {\bibinfo  {journal} {Journal of Solid State
  Chemistry}\ }\textbf {\bibinfo {volume} {145}},\ \bibinfo {pages} {33 }
  (\bibinfo {year} {1999})}\BibitemShut {NoStop}%
\bibitem [{\citenamefont {Cox}\ \emph {et~al.}(1994)\citenamefont {Cox},
  \citenamefont {Keszler},\ and\ \citenamefont {Huang}}]{Cox1994}%
  \BibitemOpen
  \bibfield  {author} {\bibinfo {author} {\bibfnamefont {J.~R.}\ \bibnamefont
  {Cox}}, \bibinfo {author} {\bibfnamefont {D.~A.}\ \bibnamefont {Keszler}},\
  and\ \bibinfo {author} {\bibfnamefont {J.}~\bibnamefont {Huang}},\ }\href
  {https://doi.org/10.1021/cm00047a021} {\bibfield  {journal} {\bibinfo
  {journal} {Chemistry of Materials}\ }\textbf {\bibinfo {volume} {6}},\
  \bibinfo {pages} {2008} (\bibinfo {year} {1994})}\BibitemShut {NoStop}%
\bibitem [{\citenamefont {Gao}\ \emph {et~al.}(2018)\citenamefont {Gao},
  \citenamefont {Xu}, \citenamefont {Tian},\ and\ \citenamefont
  {Yuan}}]{GaoJAC2018}%
  \BibitemOpen
  \bibfield  {author} {\bibinfo {author} {\bibfnamefont {Y.}~\bibnamefont
  {Gao}}, \bibinfo {author} {\bibfnamefont {L.}~\bibnamefont {Xu}}, \bibinfo
  {author} {\bibfnamefont {Z.}~\bibnamefont {Tian}},\ and\ \bibinfo {author}
  {\bibfnamefont {S.}~\bibnamefont {Yuan}},\ }\href
  {https://doi.org/https://doi.org/10.1016/j.jallcom.2018.02.110} {\bibfield
  {journal} {\bibinfo  {journal} {Journal of Alloys and Compounds}\ }\textbf
  {\bibinfo {volume} {745}},\ \bibinfo {pages} {396 } (\bibinfo {year}
  {2018})}\BibitemShut {NoStop}%
\bibitem [{sup()}]{supplementary}%
  \BibitemOpen
  \href@noop {} {\bibinfo {title} {See supplementary material for phase purity,
  crystal structure and crystal electric field model}}\BibitemShut {NoStop}%
\bibitem [{\citenamefont {Blundell}(2001)}]{BlundellMagnetism}%
  \BibitemOpen
  \bibfield  {author} {\bibinfo {author} {\bibfnamefont {S.}~\bibnamefont
  {Blundell}},\ }\href@noop {} {\emph {\bibinfo {title} {Magnetism in Condensed
  Matter}}},\ Oxford Master Series in Condensed Matter Physics\ (\bibinfo
  {publisher} {OUP Oxford},\ \bibinfo {year} {2001})\BibitemShut {NoStop}%
\bibitem [{\citenamefont {Khatua}\ \emph {et~al.}(2022)\citenamefont {Khatua},
  \citenamefont {Pregelj}, \citenamefont {Elghandour}, \citenamefont
  {Jagli\ifmmode~\check{c}\else \v{c}\fi{}ic}, \citenamefont {Klingeler},
  \citenamefont {Zorko},\ and\ \citenamefont {Khuntia}}]{Khatua_PRB2022}%
  \BibitemOpen
  \bibfield  {author} {\bibinfo {author} {\bibfnamefont {J.}~\bibnamefont
  {Khatua}}, \bibinfo {author} {\bibfnamefont {M.}~\bibnamefont {Pregelj}},
  \bibinfo {author} {\bibfnamefont {A.}~\bibnamefont {Elghandour}}, \bibinfo
  {author} {\bibfnamefont {Z.}~\bibnamefont {Jagli\ifmmode~\check{c}\else
  \v{c}\fi{}ic}}, \bibinfo {author} {\bibfnamefont {R.}~\bibnamefont
  {Klingeler}}, \bibinfo {author} {\bibfnamefont {A.}~\bibnamefont {Zorko}},\
  and\ \bibinfo {author} {\bibfnamefont {P.}~\bibnamefont {Khuntia}},\ }\href
  {https://doi.org/10.1103/PhysRevB.106.104408} {\bibfield  {journal} {\bibinfo
   {journal} {Phys. Rev. B}\ }\textbf {\bibinfo {volume} {106}},\ \bibinfo
  {pages} {104408} (\bibinfo {year} {2022})}\BibitemShut {NoStop}%
\bibitem [{\citenamefont {Stevens}(1952)}]{Stevens_1952}%
  \BibitemOpen
  \bibfield  {author} {\bibinfo {author} {\bibfnamefont {K.~W.~H.}\
  \bibnamefont {Stevens}},\ }\href {https://doi.org/10.1088/0370-1298/65/3/308}
  {\bibfield  {journal} {\bibinfo  {journal} {Proceedings of the Physical
  Society. Section A}\ }\textbf {\bibinfo {volume} {65}},\ \bibinfo {pages}
  {209} (\bibinfo {year} {1952})}\BibitemShut {NoStop}%
\bibitem [{\citenamefont {Hutchings}(1964)}]{Hutchings1964}%
  \BibitemOpen
  \bibfield  {author} {\bibinfo {author} {\bibfnamefont {M.}~\bibnamefont
  {Hutchings}}\ }(\bibinfo  {publisher} {Academic Press},\ \bibinfo {year}
  {1964})\ pp.\ \bibinfo {pages} {227--273}\BibitemShut {NoStop}%
\bibitem [{\citenamefont {Bauer}\ and\ \citenamefont
  {Rotter}(2010)}]{bauer2010magnetism}%
  \BibitemOpen
  \bibfield  {author} {\bibinfo {author} {\bibfnamefont {E.}~\bibnamefont
  {Bauer}}\ and\ \bibinfo {author} {\bibfnamefont {M.}~\bibnamefont {Rotter}},\
  }in\ \href@noop {} {\emph {\bibinfo {booktitle} {Properties and Applications
  of Complex Intermetallics}}}\ (\bibinfo  {publisher} {World Scientific},\
  \bibinfo {year} {2010})\ pp.\ \bibinfo {pages} {183--248}\BibitemShut
  {NoStop}%
\bibitem [{\citenamefont {Litvin}(2008)}]{litvin2008tables}%
  \BibitemOpen
  \bibfield  {author} {\bibinfo {author} {\bibfnamefont {D.}~\bibnamefont
  {Litvin}},\ }\href@noop {} {\bibfield  {journal} {\bibinfo  {journal} {Acta
  Crystallographica Section A: Foundations of Crystallography}\ }\textbf
  {\bibinfo {volume} {64}},\ \bibinfo {pages} {419} (\bibinfo {year}
  {2008})}\BibitemShut {NoStop}%
\bibitem [{Note1()}]{Note1}%
  \BibitemOpen
  \bibinfo {note} {We note that a related model with Heisenberg exchange
  interaction \cite {PhysRevB.104.134432} may also be relevant to our compound
  Ba$_3$Er(BO$_3$)$_3$.}\BibitemShut {Stop}%
\bibitem [{\citenamefont {Diep}\ \emph {et~al.}(1991)\citenamefont {Diep},
  \citenamefont {Debauche},\ and\ \citenamefont
  {Giacomini}}]{PhysRevB.43.8759}%
  \BibitemOpen
  \bibfield  {author} {\bibinfo {author} {\bibfnamefont {H.~T.}\ \bibnamefont
  {Diep}}, \bibinfo {author} {\bibfnamefont {M.}~\bibnamefont {Debauche}},\
  and\ \bibinfo {author} {\bibfnamefont {H.}~\bibnamefont {Giacomini}},\ }\href
  {https://doi.org/10.1103/PhysRevB.43.8759} {\bibfield  {journal} {\bibinfo
  {journal} {Phys. Rev. B}\ }\textbf {\bibinfo {volume} {43}},\ \bibinfo
  {pages} {8759} (\bibinfo {year} {1991})}\BibitemShut {NoStop}%
\bibitem [{\citenamefont {Wu}\ \emph {et~al.}(1989)\citenamefont {Wu},
  \citenamefont {Wu},\ and\ \citenamefont {Bl\"ote}}]{PhysRevLett.62.2773}%
  \BibitemOpen
  \bibfield  {author} {\bibinfo {author} {\bibfnamefont {F.~Y.}\ \bibnamefont
  {Wu}}, \bibinfo {author} {\bibfnamefont {X.~N.}\ \bibnamefont {Wu}},\ and\
  \bibinfo {author} {\bibfnamefont {H.~W.~J.}\ \bibnamefont {Bl\"ote}},\ }\href
  {https://doi.org/10.1103/PhysRevLett.62.2773} {\bibfield  {journal} {\bibinfo
   {journal} {Phys. Rev. Lett.}\ }\textbf {\bibinfo {volume} {62}},\ \bibinfo
  {pages} {2773} (\bibinfo {year} {1989})}\BibitemShut {NoStop}%
\bibitem [{\citenamefont {Metcalf}(1973)}]{METCALF19731}%
  \BibitemOpen
  \bibfield  {author} {\bibinfo {author} {\bibfnamefont {B.}~\bibnamefont
  {Metcalf}},\ }\href
  {https://doi.org/https://doi.org/10.1016/0375-9601(73)90477-5} {\bibfield
  {journal} {\bibinfo  {journal} {Physics Letters A}\ }\textbf {\bibinfo
  {volume} {45}},\ \bibinfo {pages} {1} (\bibinfo {year} {1973})}\BibitemShut
  {NoStop}%
\bibitem [{\citenamefont {Kim}(2006)}]{KIM2006245}%
  \BibitemOpen
  \bibfield  {author} {\bibinfo {author} {\bibfnamefont {S.-Y.}\ \bibnamefont
  {Kim}},\ }\href
  {https://doi.org/https://doi.org/10.1016/j.physleta.2006.05.036} {\bibfield
  {journal} {\bibinfo  {journal} {Physics Letters A}\ }\textbf {\bibinfo
  {volume} {358}},\ \bibinfo {pages} {245} (\bibinfo {year}
  {2006})}\BibitemShut {NoStop}%
\bibitem [{\citenamefont {Saito}\ and\ \citenamefont
  {Igeta}(1984)}]{doi:10.1143/JPSJ.53.3060}%
  \BibitemOpen
  \bibfield  {author} {\bibinfo {author} {\bibfnamefont {Y.}~\bibnamefont
  {Saito}}\ and\ \bibinfo {author} {\bibfnamefont {K.}~\bibnamefont {Igeta}},\
  }\href {https://doi.org/10.1143/JPSJ.53.3060} {\bibfield  {journal} {\bibinfo
   {journal} {Journal of the Physical Society of Japan}\ }\textbf {\bibinfo
  {volume} {53}},\ \bibinfo {pages} {3060} (\bibinfo {year} {1984})},\ \Eprint
  {https://arxiv.org/abs/https://doi.org/10.1143/JPSJ.53.3060}
  {https://doi.org/10.1143/JPSJ.53.3060} \BibitemShut {NoStop}%
\bibitem [{\citenamefont {Yin}\ \emph {et~al.}(2021)\citenamefont {Yin},
  \citenamefont {Li}, \citenamefont {Wang}, \citenamefont {Hu}, \citenamefont
  {Xu}, \citenamefont {Lu}, \citenamefont {Zhong}, \citenamefont {Zhao},
  \citenamefont {Zhao}, \citenamefont {Zhang}, \citenamefont {Cao},
  \citenamefont {Xu}, \citenamefont {Li}, \citenamefont {Kamiya}, \citenamefont
  {Hong},\ and\ \citenamefont {Qian}}]{PhysRevB.104.134432}%
  \BibitemOpen
  \bibfield  {author} {\bibinfo {author} {\bibfnamefont {X.}~\bibnamefont
  {Yin}}, \bibinfo {author} {\bibfnamefont {Y.}~\bibnamefont {Li}}, \bibinfo
  {author} {\bibfnamefont {G.}~\bibnamefont {Wang}}, \bibinfo {author}
  {\bibfnamefont {J.}~\bibnamefont {Hu}}, \bibinfo {author} {\bibfnamefont
  {C.}~\bibnamefont {Xu}}, \bibinfo {author} {\bibfnamefont {Q.}~\bibnamefont
  {Lu}}, \bibinfo {author} {\bibfnamefont {Y.}~\bibnamefont {Zhong}}, \bibinfo
  {author} {\bibfnamefont {J.}~\bibnamefont {Zhao}}, \bibinfo {author}
  {\bibfnamefont {X.}~\bibnamefont {Zhao}}, \bibinfo {author} {\bibfnamefont
  {Y.}~\bibnamefont {Zhang}}, \bibinfo {author} {\bibfnamefont
  {Y.}~\bibnamefont {Cao}}, \bibinfo {author} {\bibfnamefont {K.}~\bibnamefont
  {Xu}}, \bibinfo {author} {\bibfnamefont {Z.}~\bibnamefont {Li}}, \bibinfo
  {author} {\bibfnamefont {Y.}~\bibnamefont {Kamiya}}, \bibinfo {author}
  {\bibfnamefont {G.}~\bibnamefont {Hong}},\ and\ \bibinfo {author}
  {\bibfnamefont {D.}~\bibnamefont {Qian}},\ }\href
  {https://doi.org/10.1103/PhysRevB.104.134432} {\bibfield  {journal} {\bibinfo
   {journal} {Phys. Rev. B}\ }\textbf {\bibinfo {volume} {104}},\ \bibinfo
  {pages} {134432} (\bibinfo {year} {2021})}\BibitemShut {NoStop}%
\end{thebibliography}%

\end{document}